# Kinetics of Mycoprotein Production from Alternative Carbon Substrates

Mason Banks [a], Nipon Sarmah [a], Yiying He [a], Thomas Vinestock [a], Mark Taylor [b], Miao Guo [a*]

[a] Department of Engineering, Faculty of Natural Mathematical & Engineering Sciences, King's College London, Strand, London, WC2R 2LS, United Kingdom

[b] Fermentation Lead, Marlow Ingredients, Nelson Ave, Billingham, North Yorkshire, TS23 4HA, United Kingdom

[*] Corresponding author, email: miao.guo@kcl.ac.uk

## Abstract

High throughput screening was used to study of the biokinetics of *F. venenatum* A3/5 cultivation on alternative carbon substrates, including monosaccharides, disaccharides and mixtures relevant to food & beverage, dairy and agricultural waste streams. Expired functional drink from the beverage sector was also assessed as the primary carbon source for mycoprotein production. Growth data was analysed using modified single and multiphase Gompertz models for comparison of maximum specific growth rate and progression milestones across diverse growth regimes. Time-series substrate and byproduct data was analysed using comparative metrics, providing an explanatory basis for the different growth phenotypes observed. Substrate type strongly influenced the apparent carbon allocation strategies, with rapidly consumed sugars such as glucose and sucrose supporting high growth rates, low biomass yield and a high degree of fermentative byproduct formation. Fructose and xylose cultivations led to slower overall growth but higher biomass yield and lower byproduct formation. Galactose and lactose showed distinct dynamics that suggested co-existence of transport and metabolic induction limitations. In all dual-substrate systems, sequential utilisation was observed. However, metabolic inheritance and environmental shift effects were highlighted as potential kinetic limitations. These conditions exhibited stunted diauxic growth and low yield from secondary sugars, with glucose-dominated primary growth significantly reshaping secondary substrate efficiencies relative to their study *in silo*. The expired functional drink supported highly rapid growth and achieved the highest maximum specific growth rate and biomass titre of all conditions examined, alongside reduced fermentative overflow and enhanced ethanol reassimilation relative to a compositionally matched synthetic control.

# 1 Introduction

Rising global hunger and protein deficiency paired with environmental impacts and cost of food production necessitate innovative protein solutions aligned with circular economy principles to meet growing demand in a sustainable, low-cost and scalable manner [1]. Previous work demonstrated the potential of the high throughput experimental-computational workflow to provide rapid insights into the kinetics of *F. venenatum* to produce sustainable mycoprotein for food purposes from substrates emulating lignocellulosic hydrolysate [2]. Although glucose-xylose is representative of this one specific waste stream, a wider range of abundant, low-cost, carbon-rich and food-safe global side streams can also be exploited as feedstock for sustainable mycoprotein production. These may include byproducts from the dairy industry, with whey fractions derived from cheese manufacturing alone contributing between 160-170 million annual tonnes as of 2025, increasing by 2% year-on-year [3]. Whey permeate derived from cheese and milk processing contains a high fractional composition of lactose, ranging from 0.45-0.93 $g \cdot g^{-1}$ on a dry basis depending on the source and extraction method [4]. Despite its high organic content, a significant fraction of produced whey is discharged in wastewater streams, resulting in negative environmental impacts [5]. In effect, whey provides a cheap and abundant source of fermentable carbon not only in the form of lactose, but also as glucose and galactose through scaled-up feedstock hydrolysis [6]. Another resource of increasing research interest are expired fruits, juices and soft drinks from the food and beverage industry, the composition of which varies significantly depending on the resource, but contain sugars (primarily sucrose, glucose and fructose), organic acids (citric acid and malic acid), in addition to various nutrients such as vitamin C, polyphenolic compounds, and proteins/amino acids [7]. Recent works have experimentally investigated and systematically reviewed the potential of these feedstocks for various value-added products, including biofuels, biogas, microalgae technologies, bacterial cellulose, biohydrogen, bioplastics, and microbial lipids [8-12]. While these feedstocks are of increasing research interest for microbial protein production, to our knowledge, there is currently no substantial experimental literature to-date reporting the temporal utilisation efficiency or growth kinetics of *F. venenatum* for mycoprotein production on their derived substrates [13-16].

Furthermore, it was highlighted in the previous work the importance of understanding how *F. venenatum* responds to various sugars and their mixtures as the primary sources of carbon, as the kinetics and yield of sugars may vary depending on the current and previous state of the culture in mixed systems (e.g. catabolite repression effects). On this note, in the previous work the focus of analysis was limited to only three state variables (biomass, glucose and xylose concentrations) over time. However, it is important to consider from a mass balance perspective, as well as to explain the kinetics, yield and trajectory of the culture, the key byproducts produced, when and to what extent they are produced/reassimilated and from which carbon source they are primarily associated with, as these compounds may both exert and be indicative of time-varying changes in the state of the culture that may impact carbon utilisation efficiencies. This is critical in the understanding of the effectiveness of the method, as any byproduct production can help to explain substrate-specific non-idealities or limitations in the system that may violate model-based assumptions, and therefore the conclusions regarding the potential of *F. venenatum* to utilise different carbon sources.

Building on the high throughput biokinetic framework established previously, this work aims to broaden the experimental and analytical scope to understand the growth dynamics, substrate utilisation, and byproduct formation of *Fusarium venenatum* A3/5 when supplied with an array of sugars representative of broader waste streams. While the screened substrates were chosen to be representative of sugars commonly found in food- and feed-safe global waste streams, the extent to which such findings translate to real, compositionally complex feedstocks also remains unresolved. This represents a critical gap for waste-to-mycoprotein bioprocess development, as real feedstocks exhibit batch-to-batch variability, contain organic compounds beyond fermentable sugars, and may influence metabolic states in a way that cannot be inferred from sugar composition alone. Understanding how such complexity alters growth performance, carbon utilisation efficiency, and byproduct dynamics is therefore essential for both feedstock screening and process scale-up considerations.

Therefore, in addition to simple sugar formulations, an expired functional drink sample supplied by an industrial beverage manufacturer was also investigated as a representative beverage-industry waste feedstock. The sample was characterised using UHPLC to quantify major carbon components and was found to contain sucrose, glucose, fructose, and citric acid. Following pre-processing and sterilisation, the expired juice was used as the sole carbon source for *F. venenatum* batch microlitre cultivation, and time-series growth, substrate utilisation, and byproduct dynamics were quantified until depletion. To assess how growth efficiency patterns identified in synthetic screening translate to a real feedstock, a matched synthetic control containing the same four carbon substrates at equivalent concentrations was examined in parallel.

Specifically, the objectives of this work are:

1. Evaluate the kinetics of *F. venenatum* A3/5 growth on a diverse set of simple synthetic carbon substrates, including mono- and disaccharides relevant to food, beverage, dairy manufacturing and agricultural waste streams, and a real expired functional drink sample using the high throughput batch microlitre cultivation platform.
2. Quantify and compare key growth performance metrics across single- and dual-substrate systems, including maximum specific growth rate, growth milestones ($T_{10}$, $T_{50}$, $T_{90}$) and apparent biomass yields, using a phenomenological modelling approach based on single- and multi-phase modified-Gompertz functions.
3. Analyse the formation and reassimilation of major extracellular metabolic byproducts (e.g. ethanol, organic acids, and sugar alcohols) across substrates, and relate these trajectories to observed growth phases, substrate depletion patterns, and apparent carbon allocation strategies, and provide plausible mechanistic interpretations.
4. Use the comparative insights obtained to define considerations for subsequent experimental investigations to inform the development of scalable and physiologically complete models for robust and accelerated design of waste-to-mycoprotein processes.

# 2 Methodology

Several key aspects of the high throughput experimental workflow were adapted in order to address limitations found in the previous work, specifically relating to the high variability in $OD_{600}$, the low resolution of certain compounds in the UHPLC protocol, expanding the analysis to include byproducts, as well as aiming to increase the capacity and efficiency of the workflow to screen multiple conditions in a timely and resource-efficient manner. The following subsections highlight the key changes made to the initial experimental methodology outlined in the previous article [2].

## 2.1 Media preparation

Media preparation was modified to better accommodate the requirements of the high-throughput workflow. For each experiment, stock nutrient media was prepared in advance, and only the required volumes for each condition were aliquoted into sterile Falcon tubes. The appropriate mass of sugar was then added directly to each aliquot and the medium subsequently sterilised by passage through 0.22 μm syringe filters (Minisart, Sartorius).

Tween® 80 (Sigma-Aldrich) was added to all conditions at a final concentration of 0.1% (v/v). The inclusion of this surfactant was found to reduce precipitation of the mycelial mat at higher culture densities, which in turn minimised variability in $OD_{600}$ measurements. In addition, autoclave sterilisation of media and sugars was replaced entirely with syringe filtration. This adjustment was employed to mitigate precipitation of trace elements and eliminate the risk of sugar decomposition to ensure a consistent medium environment across conditions. In the case of the expired functional drink, extra steps were required in the preparation of the media. In brief, a sample was taken from a well-mixed drink and centrifuged at 3000 rpm for 10 min to remove suspended solids, the clarified liquid was decanted and passed through a 0.45 µm syringe filter (Sartorius, Minisart) to remove any remaining microparticulate matter.

Nutrient media components and trace elements were added to this liquid directly to the concentrations of the original synthetic media formulation. The pH was adjusted to 6.6 using 50 % sodium hydroxide to bring the pH within optimal range for *F. venenatum*, as preliminary tests at the original pH of the drink ($3 < pH < 4$) showed completely inhibited growth. The nutrient- and pH-adjusted juice was sterilised using a 0.22 µm syringe filter (Sartorius, Minisart) prior to inoculation with rinsed *F. venenatum* conidia. A synthetic control condition was used for comparative basis with the expired functional drink, formulated to match the initial concentrations of sugars and citric acid but added to synthetic nutrient media following the original formulation. The initial carbon composition of the expired functional drink sample was determined by UHPLC.

## 2.2 Batch microlitre fermentation of *F. venenatum A3/5*

The microlitre batch fermentation procedure generally followed the approach described in the previous work with a few key modifications. To increase throughput, a specialised high-capacity microplate incubator (INC-200D-M-120 Microplate Incubator, Cole-Parmer, USA) was employed solely for incubation, enabling multiple conditions (microplates) to be studied simultaneously. At each time point, the plates were removed from the incubator and $OD_{600}$ was measured using the microplate reader absorbance spectrometer. The lid was then removed under sterile conditions, and triplicate experimental wells were sacrificially sampled to provide three independent compositional measurements of the extracellular media. This was repeated over the fermentation time intervals until stationary phase was observed.

An additional critical adaptation was the handling of plate lids during measurements. Prior to each reading, the lid of the microplate was removed and replaced with a sterile lid, prewarmed to 28 °C inside the BSC. This step was necessary as condensation consistently accumulated on the lids during incubation, which interfered with the optical density signal producing measurement noise. The lid exchange procedure prevented artefacts caused by condensation and improved reliability of $OD_{600}$ measurements, which in general demonstrated very low well-to-well variability.

## 2.3 $OD_{600}$–CDW calibration

The experimental method for establishing $OD_{600}$–CDW correlations was refined to improve reliability. In the previous approach, cell pellets were obtained directly from the fermentation plates used for HPLC analysis. This constrained both the number of biomass samples that could be taken and the time points at which they could be collected, since sampling was determined by the replicates required for HPLC rather than by the needs of the CDW correlation. As a result, only minimal biomass was available at each time point, which was particularly limiting at low $OD_{600}$ where gravimetric measurement is less accurate.

To address this, a separate experimental plate was prepared under identical conditions solely for CDW sampling, using the same high throughput workflow as in the kinetic assays and HPLC time-series analysis. This approach provided greater flexibility in sampling design, allowing control over both the number of pooled samples collected and the $OD_{600}$ values at which they were taken. Biomass samples were collected across the full growth curve, pooled from replicate wells to overcome the small working volumes of the microplate system (0.15 mL per well), and dried to constant weight determine the CDW. The pooling strategy was adjusted to improve precision, with more replicates allocated to low $OD_{600}$ regions where biomass mass was close to the quantification limit of the balance (0.1 mg), and fewer replicates were allocated at higher $OD_{600}$, where larger biomass quantities were more readily measurable. This ensured an efficient distribution of samples across the $OD_{600}$ range to improve the accuracy of the correlation particularly in the low biomass region.

This was repeated for all conditions to obtain separate calibrations. For each condition, the $OD_{600}$ - CDW relationship was modelled using both quadratic and cubic polynomial functions as shown by Equations (1) and (2):

$$CDW = a \cdot OD_{600} + b \cdot {OD_{600}}^2 \tag{1}$$

$$CDW = a \cdot OD_{600} + b \cdot {OD_{600}}^2 + c \cdot {OD_{600}}^3 \tag{2}$$

An intercept of zero was enforced for all models to reflect the physical expectation that zero biomass represents zero $OD_{600}$, and to prevent biologically implausible negative CDW predictions at low $OD_{600}$ values. Model performance was compared using the coefficient of determination ($R^2$), root mean squared error (RMSE), and the corrected Akaike Information Criterion (AICc). Visual inspection of residuals and model behaviour across the $OD_{600}$ range was also used to guide selection.

While cubic models frequently yielded the lowest RMSE, particularly by better capturing the early, near-linear portion of the calibration curve before $OD_{600}$ saturation, quadratic functions were deliberately selected in cases where cubic fits exhibited signs of overfitting or unstable extrapolation at high $OD_{600}$. This choice prioritised generalisability and reduced uncertainty in the high-OD region, which is especially relevant when applying the calibration to experimental data approaching the upper measurement range where the Beer-Lambert assumption no longer holds.

Across all tested conditions, model fits were strong ($R^2 \geq 0.86$), with RMSE values generally below 0.45 $g \cdot L^{-1}$. In most single sugar conditions, cubic models provided marginally better statistical performance. However, quadratic fits were selected for glucose, xylose, fructose, lactose, galactose, and both glucose–galactose and glucose–xylose mixtures to mitigate high overfitting. Cubic fits were retained where data supported stable behaviour across the $OD_{600}$ range, such as for sucrose (RMSE = 0.106 $g \cdot L^{-1}$, $R^2$ = 0.998). The lowest calibration accuracy was observed for lactose ($R^2$ = 0.859).

To give a measure of the uncertainty in biomass estimates derived from $OD_{600}$ measurements, variability in $OD_{600}$ replicates was propagated through the selected calibration function to provide an upper and lower uncertainty bounds on CDW estimates (SI). These calibrations form a key component of the high-throughput workflow, allowing CDW to be estimated directly from $OD_{600}$ readings under a variety of conditions without reliance on traditional dilution-based methods unsuitable for filamentous cultures. Full parameter sets and summary statistics for each condition are provided in Supplementary Information 2.

## 2.4 UHPLC analysis of sugars, sugar alcohols, organic acids, and ethanol

In the previous work, the HPLC protocol was limited by insufficient separation of closely eluting compounds, most notably the significant overlap between phosphate and glucose at low concentrations. To overcome this, the method was refined by adjusting the oven temperature, mobile phase acidity, flow rate, and injection volume. These modifications improved peak resolution, reduced overlap, and extended the protocol to enable quantification of fermentation byproducts, thereby providing a more complete analysis of the system.

The same Aminex HPX-87H column was used but two distinct operating protocols were employed in this study.

### 2.4.1 Protocol 1: Quantification of non-heat sensitive media

The operating conditions for protocol 1 were 30 °C oven temperature, 0.4 mL/min flow rate, 1 µL injection volume, and 0.001 M $H_2SO_4$ as the mobile phase. These settings provided reliable separation of sugar substrates while simultaneously resolving major fermentation byproducts. The reduced acidity also caused the phosphate peak to elute earlier relative to the sugar peaks, alleviating the interference observed in the previous work.

### 2.4.2 Protocol 2: Quantification of sucrose-containing media

For conditions containing sucrose, additional modifications were required to prevent inversion of sucrose into glucose and fructose within the column. In this case, the oven temperature was lowered to 15 °C and the mobile phase acidity reduced to 0.0005 M $H_2SO_4$. While this protocol theoretically permits analysis of all conditions, its use was limited in practice. Operation below ambient temperature caused condensation within the column oven, and the acidity (approaching the operating pH limit of the $H^+$ cation-exchange resin) risked long-term deterioration of column performance. In addition, the combination of reduced temperature and flow rate resulted in elevated backpressure (up to 103 bar), approaching the maximum recommended operating pressure. For these reasons, Protocol 2 was reserved exclusively for sucrose-containing conditions where accurate quantification was essential.

For both protocols, external calibration curves were generated from reference standards for each analyte. Full calibration data and compound-specific retention times are provided in Supplementary Information 1.

### 2.4.3 Expired functional drink characterisation

The initial carbon compositions measured using this protocol for the expired functional drink and its synthetic control condition are provided in Table 1. It should be noted that in a separate work by He *et al.* (2025) investigating the biomethane potentials of functional drinks, compositional characterisation of the same drink employed in this study quantified minor concentrations of oxalic acid (<0.1 g/L), malic acid (<0.35 g/L), in addition to free amino acids alanine (~ 4.6 mmol/L), aspartic acid (1.3 mmol/L) and sum of others (~0.9 mmol/L) [17]. However, these compounds were undetectable with the protocol employed in this work and are therefore untracked in the time-series analysis.

*Table 1. Initial composition of expired function drink measured using HPLC Protocol 2.*

| | **Concentration** (g/L) | |
|---|---|---|
| **Compound** | **Expired Functional Drink** | **Synthetic Control** |
| Citric Acid | 7.04 ± 0.01 | 6.8 ± 0.03 |
| Fructose | 13.48 ± 0.06 | 12.85 ± 0.05 |
| Glucose | 13.24 ± 0.03 | 13.24 ± 0.1 |
| Sucrose | 3.06 ± 0.01 | 3.1 ± 0.03 |

## 2.5 Data analysis framework

The primary modelling approach in this work was data-driven and phenomenological metrics were derived as follows to allow for fair cross-comparison between conditions.

### 2.5.1 Modified Gompertz/multi-Gompertz growth curve analysis

The time evolution of biomass concentration, $X(t)$, was described using the Zwietering-modified Gompertz model, which is widely applied in microbial growth analysis due to its ability to capture microbial growth behaviour with a small set of interpretable parameters [18-20]. A single Gompertz component $g$ is defined as:

$$g(t, a, \kappa, \lambda) = A \cdot exp\left[-exp\left(\frac{\kappa \cdot e}{A}(\lambda - t) + 1\right)\right] \quad (3)$$

Where $A$ is the carrying capacity ($g \cdot L^{-1}$), $\kappa$ is the rate scale parameter ($g \cdot L^{-1} \cdot h^{-1}$), $\lambda$ is the lag phase duration (h), $e$ is Euler's number (≈2.718), and $t$ is time (h).

As some datasets demonstrated atypical growth behaviours, a multi-phase modified Gompertz model was tested across the datasets, where the multi-phase model is given by Equation (4).

$$X(t) = X_0 + \sum_{i=1}^{N} g(t, a, \kappa, \lambda) = X_0 + \sum_{i=1}^{N} A_i \cdot exp\left[-exp\left(\frac{\kappa_i \cdot e}{A_i}(\lambda_i - t) + 1\right)\right] \qquad (4)$$

Where $X_0$ is the baseline biomass concentration (g·L$^{-1}$) and $i$ indicates the phase-specific parameters.

Parameters for both the single- and multi-phase Gompertz models were estimated by nonlinear least squares regression using the same differential evolution optimisation algorithm and objective formulation specified in Banks *et al* (2024). Selection of the number of Gompertz components was based on the corrected Akaike Information Criterion (AICc), with preference given to the simpler model (lower AICc value). The $R^2$ and $RMSE$ were determined for each model to evaluate goodness-of-fit. The full sets of estimated Gompertz parameters and summary statistics are provided in Supplementary Information 3.

### 2.5.2 Maximum biomass concentration ($X_{max}$)

The maximum observed biomass concentration, $X_{max}$ (g·L$^{-1}$), was defined as the maximum mean CDW measured over the course of the experiment occurring at time $t_{X,max}$ (h).

### 2.5.3 Maximum specific growth rate ($\mu_m$)

To estimate the $\mu_m$ (h$^{-1}$), the time derivative of the fitted Gompertz function was normalised by the corresponding biomass to obtain the instantaneous specific growth rate. The maximum specific growth rate was then defined as the maximum value of the specific growth rate occurring after the first modelled lag phase (Equation (5)).

$$\mu_m = \max_{t > \lambda}\left[\frac{1}{X(t)}\frac{dX(t)}{dt}\right] \qquad (5)$$

This constraint was applied because derivative-based or graphically derived estimates in regions of very low biomass are highly sensitive to numerical noise and sparse sampling and may produce artificially inflated $\mu$ values. This issue was particularly pronounced when estimating $\mu_m$ from log-transformed CDW data, where small initial biomass values resulted in unrealistically high apparent growth rates.

### 2.5.4 Diauxic plateau onset ($t_{DP}$) and duration ($\lambda_{DP}$)

The duration of the diauxic plateau region, $t_{DP}$ (h), between two Gompertz growth phases is defined as the time between the onset of the plateau, $\lambda_{DP}$ (h), and the start of the subsequent growth phase, $\lambda_2$ (h):

$$DP = \lambda_2 - \lambda_{DP} \tag{6}$$

Where $\lambda_{DP}$ is defined as the intersection between the horizontal line corresponding to the carrying capacity of the first phase ($y = A_1$), and the tangent of this phase at its inflection point:

$$\frac{d^2 g_1(t)}{dt^2} = 0 \tag{7}$$

### 2.5.5 Growth milestones ($T_{10}$, $T_{50}$, $T_{90}$)

To assess the relative growth progress of each condition, the times to reach 10%, 50%, and 90% of the maximum experimental biomass were obtained directly from the fitted Gompertz growth curve. For each fitted trajectory $X(t)$, $T_{10}$, $T_{50}$, and $T_{90}$ were defined as the times at which biomass reached 10%, 50%, and 90% of the total biomass increase, respectively:

$$X(T_f) = X_0 + f\,(X_{max} - X_0), \qquad f \in (0.1, 0.5, 0.9) \tag{8}$$

where $X_0$ is the baseline biomass concentration and $X_{max}$ is the maximum observed biomass.

### 2.5.6 Substrate utilisation rate ($qS_{avg}$)

Substrate utilisation rates, $qS_{avg}$ ($g \cdot L^{-1} \cdot h^{-1}$), were calculated as average depletion rates between 90% and 10% of the initial substrate concentration ($S_{\sim 90}$ to $S_{\sim 10}$), as shown in Equation (9). The corresponding time points were identified directly from the experimental substrate profiles as the sampling points at which the substrate concentration was closest to these levels. This interval was selected to capture the period of sustained substrate consumption while minimising the influence of early adaptation effects and late-stage depletion near the detection limit.

$$qS_{avg} = \frac{S_{\sim 90} - S_{\sim 10}}{t_{S\sim 10} - t_{S\sim 90}} \tag{9}$$

2.5.7 Biomass yield ($Y_{X/S}$)

Biomass yields, $Y_{X/S}$ (g biomass/g substrate), were calculated as apparent yield coefficients defined as the ratio of biomass formed to substrate consumed within a specified time window:

$$Y_{X/S} = \frac{\Delta X}{\Delta S} \tag{10}$$

Where $\Delta X$ is the change in biomass concentration and $\Delta S$ is the corresponding substrate consumption in the same interval.

Total biomass yield ($Y_{X/S}$) was calculated from the start of the experiment to the time of substrate depletion, $t_{S\sim 0}$ (Equation (11)).

$$Y_{X/S} = \frac{X(t_{S\sim 0}) - X_{in}}{S_{in} - S(t_{S\sim 0})} \tag{11}$$

Where $X_{in}$ and $S_{in}$ are the initial biomass and substrate concentrations respectively.

For dual-substrate conditions, apparent yield efficiencies were calculated separately for the primary ($S_1$) substrate (i.e. glucose) and secondary ($S_2$) substrates (i.e. xylose or galactose) using phase-specific windows defined by their availability, according to Equations (12) and (13).

$$Y_{X/S1} = \frac{X(t_{S1\sim 0}) - X_{in}}{S_{1,in} - S(t_{S1\sim 0})} \tag{12}$$

$$Y_{X/S2} = \frac{X(t_{S2\sim 0}) - X(t_{S1\sim 0})}{S_2(t_{S1\sim 0}) - S_2(t_{S2\sim 0})} \tag{13}$$

Given the presence of citric acid in the expired functional drink and its differing carbon content on a mass basis relative to sugars, overall apparent biomass yields were calculated on a carbon-molar basis as opposed to a mass basis. The total initial carbon supplied by sugars and citric acid was therefore used as the denominator in the yield calculation. Biomass concentrations were converted from g·L$^{-1}$ to carbon-molar units using the carbon mass fraction of *F. venenatum* biomass (40.05% C per unit dry weight), determined by elemental analysis of industrial Quorn mycoprotein [21] . This enabled calculation of the overall apparent biomass yield on total supplied carbon according to Equation (14).

$$Y_{\frac{X}{S+CA}} = \frac{\Delta nC_X}{\Delta nC_{S+CA}} = \frac{nC_{X,out} - nC_{X,in}}{nC_{S+CA,in} - nC_{S+CA,out}} \quad (14)$$

Where $\Delta nC_X$ is the carbon incorporated into biomass (C-mol) and $\Delta nC_{C+SA}$ is the total carbon supplied by citric acid and sugars (C-mol) from the start to the end of the time-series.

2.5.8 Maximum product concentration ($P_{max}$)

For each measured product $P$, the maximum observed concentration $P_{max}$ (g·L$^{-1}$) and the time at which it occurred, $t_{P,max}$ (h), were reported.

2.5.9 Maximum product formation rate ($qP_{max}$)

The maximum product formation rate, $qP_{max}$ (g·L$^{-1}$·h$^{-1}$), was estimated directly from experimental product concentration profiles using finite differences as a measure of production intensity. Product formation rates were calculated only over the formation phase, from the onset of detectable product formation $t_{P,on}$ (h) to the time of maximum product concentration $t_{P,max}$ (h). Local product formation rates were computed using a sliding window of three consecutive time points, and the maximum calculated slope was reported as $qP_{max}$.

$$qP_i = \frac{P_{i+1} - P_{i-1}}{t_{i+1} - t_{i-1}} \text{ for } i \in [\, i_{on} + 1,\, i_{max} - 1\,] \quad (14)$$

$$qP_{max} = \max\,(qP_i) \quad (15)$$

Where $P_i$ and $t_i$ denote the measured product concentration and time at sampling point $i$, $i_{on}$ corresponds to $t_{P,on}$, and $i_{max}$ corresponds to $t_{P,max}$ .

2.5.10 Product formation yield ($Y_{P/S}$)

The gross apparent product yields were calculated from the start time of product formation ($t_{P,on}$) to $t_{P,max}$:

$$Y_{P/S} = \frac{P_{max} - P_{on}}{S(t_{P,on}) - S(t_{P,max})} \tag{16}$$

Where $t_{P,on}$ is defined as the first time a product was reliably detected, i.e.

$$t_{P,on} = \min\left[t_i : P(t_i) \geq \text{LOQ for 2 consecutive samples}\right] \tag{17}$$

### 2.5.11 Average product reassimilation rate ($qP_{reassim}$)

The average rate of product reassimilation $qP_{reassim}$ ($g \cdot L^{-1} \cdot h^{-1}$) was calculated using finite differences, from the time whereat sustained re-uptake began ($t_{P,reassim,on}$) to the time of final measurable concentration ($t_{P,final}$):

$$qP_{reassim} = \frac{P_{reassim,on} - P_{final}}{t_{P,reassim,on} - t_{P,final}} \tag{18}$$

# 3 Results

## 3.1 Single substrate kinetics

Figure 1 shows the time series experimental results observed for *F. venenatum* growth on 30g·L$^{-1}$ glucose (A); xylose (B); fructose (C); sucrose (D); lactose (F); and 15g·L$^{-1}$ galactose (E). The biomass dynamics, substrate utilisation kinetics, and byproduct formation / reassimilation are discussed in the following subsections.

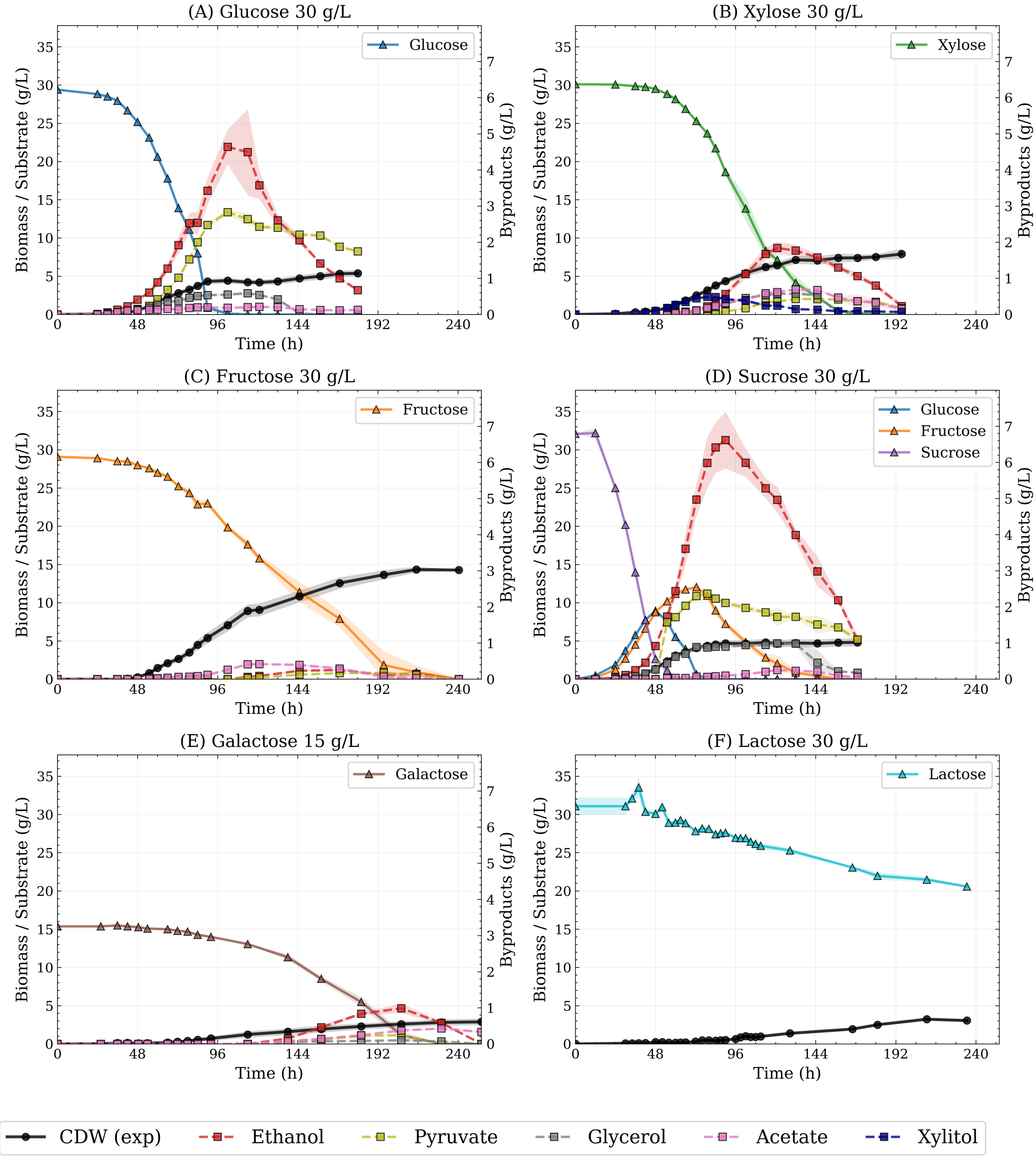

*Figure 1. Time-series experimental profiles of F. venenatum growth and byproduct formation / reassimilation on various synthetic single substrates. (A) Glucose 30g·L$^{-1}$; (B) Xylose 30g·L$^{-1}$; (C) Fructose 30g·L$^{-1}$; (D) Sucrose 30g·L$^{-1}$; (E) Galactose 15g·L$^{-1}$; (F) Lactose 30g·L$^{-1}$. Points and shaded bands represent the experimental mean and standard deviation respectively. The mean trend is indicated by a line.*

### 3.1.1 Growth dynamics across single sugars

For the single substrate conditions, the modified Gompertz models demonstrated high accuracy, with $R^2$ near unity and low RMSE for all datasets demonstrating excellent goodness-of-fit. The appropriate model complexity was well justified by the AICc, with ΔAICc ranging between 11 and 19 when compared with the runner-up model (Table 2). Figure 2 presents the Gompertz model fits against experimental data, and Table 3 presents the key Gompertz derived metrics discussed herein.

*Table 2. Gompertz model selection and goodness-of-fit metrics for biomass growth under single substrate conditions. The preferred number of growth phases was selected using AICc, with ΔAICc indicating separation from the next-best model. RMSE and $R^2$ quantify fit accuracy.*

| **Condition** | **Phases** | **AICc** | **AICc (runner-up)** | **ΔAICc** | **RMSE** (g·L$^{-1}$) | $R^2$ |
|---|---|---|---|---|---|---|
| **Glucose 30 g·L$^{-1}$** | 2 | -60 | -48 | 12 | 0.17 | 0.99 |
| **Xylose 30 g·L$^{-1}$** | 1 | -82 | -71 | 11 | 0.12 | 1 |
| **Fructose 30 g·L$^{-1}$** | 2 | -47 | -28 | 19 | 0.21 | 1 |
| **Sucrose 30 g·L$^{-1}$** | 1 | -93 | -77 | 16 | 0.1 | 1 |
| **Galactose 15 g·L$^{-1}$** | 2 | -103 | -92 | 11 | 0.04 | 1 |

Glucose-supported growth was best characterised by two phases, with a maximum biomass concentration of 5.38 ± 0.42 g·L$^{-1}$. The first fitted phase followed a standard sigmoidal profile, with an estimated maximum specific growth rate of 0.13 h$^{-1}$. The growth progress was relatively rapid during this first phase, with a lag time of 44.6 h, and $T_{10}$ and $T_{50}$ separated by only ~22 h. The first phase was dominant relative to the second, accounting for ~81% of the total biomass formed and $\lambda_2$ occurring only ~15 h prior to the $T_{90}$. Furthermore, although the model fit was reasonable for this condition, it should be noted that experimentally, there appears to be a small but significant decline in biomass during the time between phases that cannot be captured by the Gompertz model due to its monotonic structure. Instead, a diauxic plateau of 55.5 h was predicted to compensate for this, and as such, the model slightly underpredicts the contribution of the two phases due to this averaging of growth components.

Xylose exhibited slower and more temporally distributed growth, achieving a relatively high maximum biomass concentration of 7.91 ± 0.69 g·L$^{-1}$. The growth curve was best described by a single phase, with relatively well-balanced time intervals between growth milestones when compared with multiphasic conditions. The biomass underwent a slightly longer lag time than glucose (by ~5 h) before entering the exponential growth phase with a low maximum specific growth rate of 0.09 $h^{-1}$.

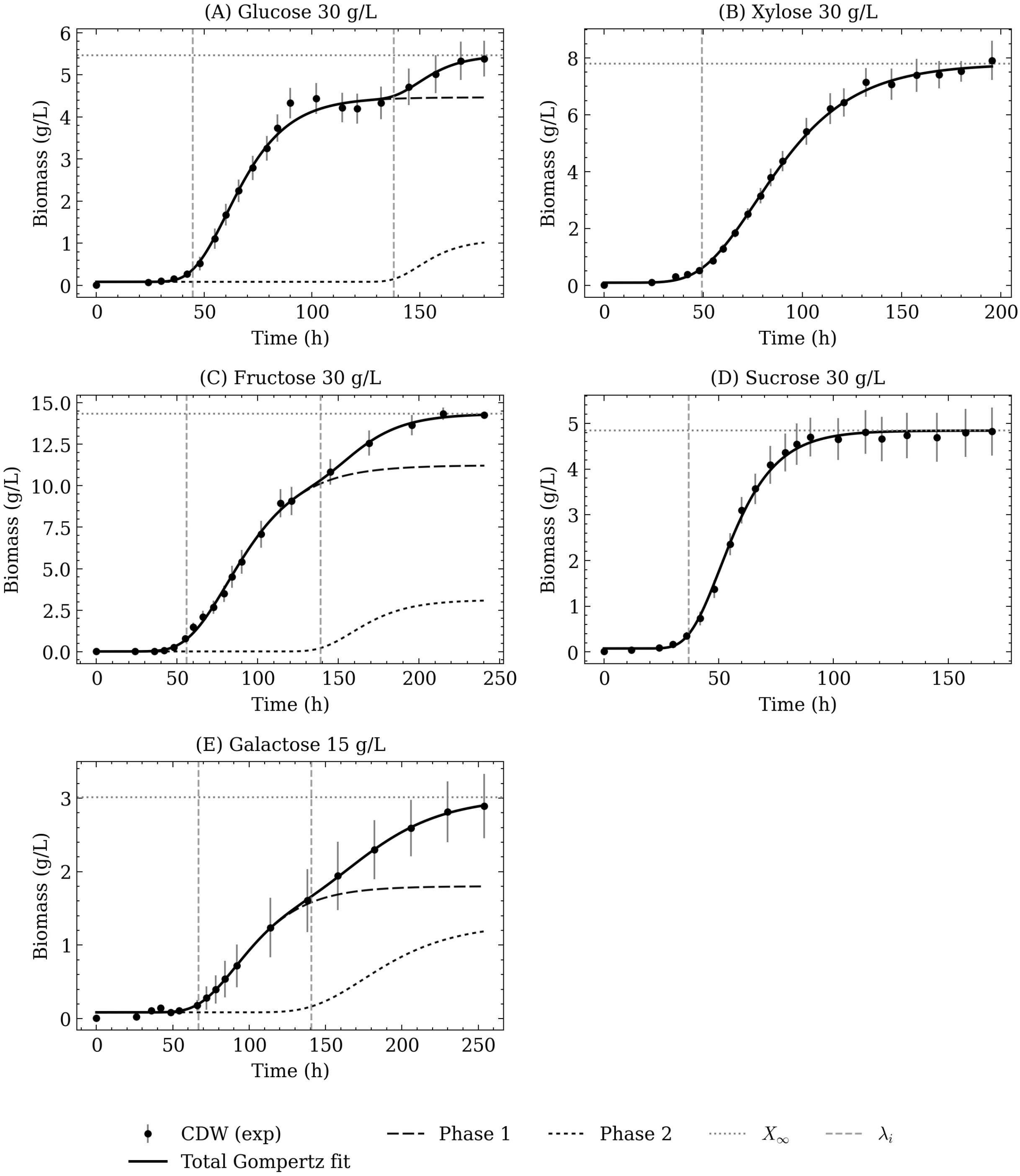

*Figure 2. Model simulated best fit modified Gompertz functions plotted against experimental CDW data for single synthetic substrates. (A) Glucose 30g·L$^{-1}$; (B) Xylose 30g·L$^{-1}$; (C) Fructose 30g·L$^{-1}$; (D) Sucrose 30g·L$^{-1}$; (E) Galactose 30g·L$^{-1}$. Black circles and vertical bars represent the mean and uncertainty of the CDW respectively.*

*Table 3. Summary of growth performance metrics for single substrates. Experimental values are reported as mean ± standard deviation. * Gompertz derived metrics not determined for lactose condition due to poor model fit.*

| **Metric** | **Glucose (30g·L$^{-1}$)** | **Xylose (30g·L$^{-1}$)** | **Fructose (30g·L$^{-1}$)** | **Sucrose (30g·L$^{-1}$)** | **Galactose (15g·L$^{-1}$)** | **Lactose (30g·L$^{-1}$) *** |
|---|---|---|---|---|---|---|
| $\boldsymbol{X_{max}}$ (g·L$^{-1}$) | 5.38 ±0.42 | 7.91 ±0.69 | 14.34 ±0.07 | 4.82 ±0.52 | 2.89 ±0.44 | 3.24 ±0.17 |
| $\boldsymbol{t_{X_{max}}}$ (h) | 180 | 195.5 | 215 | 169 | 254 | 210.5 |
| $\boldsymbol{\mu_m}$ (h$^{-1}$) | 0.13 | 0.09 | 0.1 | 0.16 | 0.06 | - |
| $\boldsymbol{\lambda_1}$ (h) | 44.6 | 49.3 | 55.8 | 37 | 66.8 | - |
| $\boldsymbol{\lambda_2}$ (h) | 138 | - | 138.9 | - | 140.8 | - |
| $\boldsymbol{\lambda_{DP}}$ (h) | 87.5 | - | 124.9 | - | 134 | - |
| $\boldsymbol{DP}$ (h) | 50.5 | - | 14 | - | 6.7 | - |
| $\boldsymbol{T_{10}}$ (h) | 48.6 | 54 | 62.9 | 39.2 | 76.8 | - |
| $\boldsymbol{T_{50}}$ (h) | 71.2 | 87 | 101.6 | 55.5 | 129 | - |
| $\boldsymbol{T_{90}}$ (h) | 150.6 | 141.7 | 173.6 | 80.8 | 206.7 | - |

Fructose led to the highest biomass concentration of all single substrates, reaching 14.34 ± 0.07 g·L$^{-1}$, but did so over the longest required incubation time (240.5 h) and was best fit by a two-phase model structure. The estimated maximum specific growth rate was 0.1 h$^{-1}$, and growth milestones were delayed relative to glucose, sucrose and xylose, with a particularly protracted period (72 h) of slow growth between $T_{50}$ and $T_{90}$, wherein the model-predicted second phase begins centrally at $\lambda_2$ (~139 h). The estimated transition time ($DP$) between phases was short relative to glucose, requiring only 14 h.

Sucrose supported the most rapid accumulation of biomass, with the earliest predicted lag phase duration (37 h) and the highest maximum specific growth rate (0.16 h$^{-1}$), closest to that reported for industrial mycoprotein production from glucose (0.17 - 0.19 h$^{-1}$) [22]. The growth milestones occurred early and with the smallest separation, with $T_{10}$, $T_{50}$, and $T_{90}$ occurring at 39.2 h, 55.5 h, and 80.8 h, respectively. The maximum biomass concentration achieved was relatively low (4.82 ± 0.52 g·L$^{-1}$), comparable to glucose and substantially lower than fructose or xylose. A single Gompertz component was sufficient to describe the data, indicative of an intense growth regime with relatively early cessation.

Galactose-supported growth was very slow and limited in extent, exhibiting the lowest maximum specific growth rate (0.06 $h^{-1}$) and the latest progression through growth milestones, with $T_{10}$, $T_{50}$, and $T_{90}$ occurring at 76.8 h, 129.0 h, and 206.7 h, respectively. The maximum biomass concentration achieved was also the lowest of all conditions (2.89 ± 0.44 $g \cdot L^{-1}$). Although a two-phase Gompertz model was selected, both phases were slow and contributed to a prolonged, low-intensity growth trajectory rather than a clear shift between rapid and slow regimes as in the case of glucose, with only 58% of the total biomass accounted for by the first phase.

Lactose-supported growth differed structurally from the other substrates and did not abide to a sigmoidal growth trajectory typical of carbon-limited cultures. A maximum biomass concentration of 3.24 ± 0.17 $g \cdot L^{-1}$ was achieved, only slightly higher than that achieved by its galactose moiety studied separately. Growth progressed slowly and approximately linearly, and as such, lactose was excluded from comparative Gompertz-based analysis and is interpreted as a kinetically unique growth regime outside the scope of phenomenological growth models applied here.

### 3.1.2 Substrate utilisation kinetics

Substrate utilisation kinetics also differed across the tested carbon sources, revealing trade-offs between uptake rate, duration of substrate availability, and apparent biomass yields. Table 4 shows the key metrics of substrate utilisation kinetics for single substrates.

*Table 4. Summary of substrate utilisation kinetics for single substrates. Experimental values are reported as mean ± standard deviation. † For sucrose, substrate uptake rates and yields were calculated on a hexose-equivalent basis. *Lactose was not fully consumed in the experimental timeframe.*

| **Metric** | **Glucose ($30g \cdot L^{-1}$)** | **Xylose ($30g \cdot L^{-1}$)** | **Fructose ($30g \cdot L^{-1}$)** | **Sucrose ($30g \cdot L^{-1}$) †** | **Galactose ($15g \cdot L^{-1}$)** | **Lactose ($30g \cdot L^{-1}$)** |
|---|---|---|---|---|---|---|
| $\boldsymbol{S_{in}}$ ($g \cdot L^{-1}$) | 29.4 ±0.17 | 30.11 ±0.05 | 29.09 ±0.07 | 32.08 ±0.30 | 15.37 ±0.04 | 31.09 ±1.14 |
| $\boldsymbol{t_{S\approx 0}}$ (h) | 102 | 180 | 240.5 | 157.5 | 230 | -* |
| $\boldsymbol{qS_{avg}}$ ($g \cdot L^{-1} \cdot h^{-1}$) | 0.538 | 0.312 | 0.186 | 0.281 | 0.113 | 0.057 |
| $\boldsymbol{Y_{X/S}}$ ($g \cdot g^{-1}$) | 0.15 | 0.25 | 0.49 | 0.15 | 0.18 | 0.29 |

Glucose was consumed most rapidly of all substrates, declining from an initial concentration of 29.4 $g \cdot L^{-1}$ to near depletion by 102 h at the highest observed uptake rate (0.538 $g \cdot L^{-1} \cdot h^{-1}$).

The biomass yield on glucose was relatively low (0.15 g·g$^{-1}$), indicating that a substantial fraction of assimilated carbon was not converted to biomass. Notably, glucose depletion occurred shortly after the Gompertz predicted transition to diauxic plateau ($\lambda_{DP}$ = 87.5 h), indicating that the first phase was primarily governed by glucose limitation. The combination of fast uptake, early depletion, and low yield is consistent with a growth strategy prioritising rapid carbon flux over efficient biomass formation.

Xylose exhibited a slower utilisation profile, requiring approximately 180 h to reach near depletion, with an average uptake rate of 0.312 g·L$^{-1}$·h$^{-1}$. In contrast to glucose, xylose supported a higher biomass yield (0.25 g·g$^{-1}$). Notably, the $T_{90}$ (141.7 h) was reached long before complete xylose depletion (180 h), indicating that xylose availability alone did not dictate growth slowdown, and that utilisation efficiency declined during the late stage of cultivation. g·g$^{-1}$

Fructose was utilised at a relatively slow rate of 0.186 g·L$^{-1}$·h$^{-1}$ and remained detectable up to 240.5 h. However, this prolonged utilisation was associated with an exceptionally high apparent biomass yield (0.49 g·g$^{-1}$), more than threefold higher than that observed for glucose. This indicates that fructose-supported growth favoured sustained, efficient carbon incorporation into biomass rather than rapid flux through central metabolism in contrast to glucose, despite their molecular similarity.

Sucrose presented as a special case relative to the other substrates studied, as it was hydrolysed rapidly (~0.82 g·L$^{-1}$·h$^{-1}$) by *F. venenatum* early in the culture, releasing extracellular glucose and fructose which were then sequentially utilised. This is likely a result of invertase secretion observed in many fungal species [23]. Total hexose reached effective depletion by 157.5 h corresponding to an average uptake rate of 0.281 g·L$^{-1}$·h$^{-1}$, representing intermediate substrate kinetics between rapid glucose and slower fructose utilisation. The apparent biomass yield on sucrose (0.15 g·g$^{-1}$) was identical to that observed for glucose. This similarity in yield indicates that the presence of a glucose moiety within sucrose strongly influenced overall carbon utilisation efficiency. This is supported by the observation that fructose was utilised very slowly post-glucose depletion and contributed only marginally to the formation of new biomass.

Galactose was utilised slowest relative to the other monosaccharides, with low overall efficiency. Despite being supplied at a lower initial concentration (15.4 g·L$^{-1}$), galactose required approximately 230 h to reach near depletion. The substrate concentration only began to decline significantly after 72 h, consistent with the $\lambda_1$ estimated by the Gompertz model. The subsequent rate of utilisation was the lowest of the monosaccharides at 0.113 $g\cdot L^{-1}\cdot h^{-1}$, and the apparent biomass yield on galactose was only slightly higher than glucose (0.18 $g\cdot g^{-1}$).

Lactose represented an extreme case of limited carbon conversion, with only 34% of the supplied sugar consumed over 234.5 h. The average uptake rate was also the lowest observed overall (0.057 $g\cdot L^{-1}\cdot h^{-1}$). On the other hand, the biomass yield calculated over the utilisation window was relatively high (0.29 $g\cdot g^{-1}$), suggesting slow but efficient assimilation.

### 3.1.3 Byproduct formation and reassimilation

Byproduct formation and reassimilation patterns demonstrated considerable variations and similarities with regards to formation intensity, onset and reassimilation timings, and apparent yields. By assessing these trends, discrepancies in growth behaviours and substrate utilisation efficiencies could be effectively resolved, allowing for deeper insights into carbon allocation strategies and potential unmeasured kinetic limitations. In general, the most prominent byproduct across conditions was ethanol, with secondary metabolites including pyruvate, acetate, glycerol, and xylitol, the latter of which was only measured during xylose utilisation. Key metrics are presented in Table 5 with emphasis on ethanol as the predominant byproduct. Comprehensive metrics are presented in Supplementary Information 4.

Glucose and sucrose cultures exhibited similar byproduct trajectories, producing the highest concentrations of measured ethanol, pyruvate and glycerol across conditions. Production onset occurred early in both cases, with detectable quantities appearing within 7-9 h prior to the estimated lag phase duration. Formation occurred rapidly and was persistent throughout the exponential growth phase until glucose reached near depletion (102 h and 90 h for glucose and sucrose conditions respectively), indicating that glucose was the primary driver of growth-associated byproduct formation. However, the 2-fold discrepancy in the yield of ethanol between glucose and sucrose may indicate that some portion of fructose was co-utilised to this end. Indeed, the peak fructose concentration measured after its rise from hydrolytic release was only 71% of the maximum expected from complete sucrose inversion. In both cases, ethanol was most rapidly reassimilated post-glucose depletion, achieving ~83-85% fractional reassimilation. Pyruvate was reassimilated more gradually and exhibited a low fractional reassimilation (~38-54%). Glycerol presented as an interesting case, persisting after glucose depletion  around its maximum concentration until ~121 h, after which it was rapidly reassimilated. Although produced at low concentrations, acetate demonstrated slightly different behaviour between the two conditions, with acetate production persisting after 132 h in the case of sucrose, while in the case of glucose, acetate began to decline after this time.

*Table 5. Byproduct formation and reassimilation metrics for single substrates. Experimental values are reported as mean ± standard deviation. † For sucrose, ethanol yield was calculated on a hexose-equivalent basis. *Insufficient finite slopes to quantify. n.d. = not detected; EtOH = ethanol; Pyr = pyruvate; Gly = glycerol; Ace = acetate.*

| Metric | Glucose ($30 g·L^{-1}$) | Xylose ($30 g·L^{-1}$) | Fructose ($30 g·L^{-1}$) | Sucrose ($30 g·L^{-1}$) † | Galactose ($15 g·L^{-1}$) | Lactose ($30 g·L^{-1}$) |
|---|---|---|---|---|---|---|
| $\boldsymbol{EtOH_{max}}$ ($g·L^{-1}$) | 4.64 ±0.49 | 1.84 ±0.17 | 0.25 ±0.02 | 6.62 ±0.79 | 0.99 ±0.13 | n.d. |
| $\boldsymbol{t_{EtOH_{max}}}$ (h) | 102 | 121 | 169 | 90 | 206 | - |
| $\boldsymbol{qP_{EtOH,max}}$ ($g·L^{-1}·h^{-1}$) | 0.117 | 0.046 | < 0.01 | 0.203 | 0.012 | - |
| $\boldsymbol{Y_{EtOH/S}}$ ($g·g^{-1}$) | 0.16 | 0.1 | 0.02 | 0.32 | 0.08 | - |
| $\boldsymbol{qP_{reassim,EtOH}}$ ($g·L^{-1}·h^{-1}$) | 0.051 | 0.022 | -* | 0.07 | -* | - |
| $\boldsymbol{Pyr_{max}}$ ($g·L^{-1}$) | 2.83 ±0.08 | 0.43 ±0.02 | 0.18 ±0.01 | 2.37 ±0.18 | 0.25 ±0.03 | n.d. |
| $\boldsymbol{Gly_{max}}$ ($g·L^{-1}$) | 0.59 ±0.02 | 0.57 ±0.03 | n.d. | 0.99 ±0.09 | 0.10 ±0.01 | n.d. |
| $\boldsymbol{Ace_{max}}$ ($g·L^{-1}$) | 0.21 ±0.02 | 0.69 ±0.02 | 0.42 ±0.02 | 0.25 ±0.03 | 0.43 ±0.04 | n.d. |

These observations align well with the growth dynamics and substrate utilisation efficiencies of glucose and sucrose. It appears that in both cases, the higher growth and glucose uptake rates may have exceeded the anabolic capacity of the system, resulting in a high allocation of carbon to byproduct formation and a subsequent reduction in the overall biomass yield. Furthermore, in the case of glucose, the slower second Gompertz phase appears to correspond to delayed biomass formation from the reuptake of byproducts, while in the case of sucrose, co-consumption of fructose and byproducts occurred during a constant stationary phase.

In contrast, xylose supported delayed and moderate byproduct formation, with an ethanol yield of 0.1 $g·g^{-1}$. While still growth-associated, the onset of all measured byproducts (besides xylitol) occurred during mid-exponential growth (close to $T_{50}$) and were produced primarily during the most rapid period of xylose uptake, reaching their maximum concentrations between 121 and 134 h. While glycerol and pyruvate trajectories mirrored the overall dynamics of those conditions, the maximum pyruvate concentration was significantly lower (0.43 ± 0.02 $g \cdot L^{-1}$). This indicates that respiratory metabolism was insufficient to fully accommodate xylose-derived carbon flux during this period, albeit to a lesser extent than glucose/sucrose. Interestingly, xylitol was the earliest detected byproduct and was produced from the onset of exponential growth until $T_{50}$, whereafter it steadily declined to trace amounts around $T_{90}$. Similarly to glucose and sucrose, ethanol was reassimilated most rapidly, achieving ~88% completion by 195.5 h. Critically, the reuptake of fermentative byproducts began as the residual xylose concentration declined from 24-14% of its initial concentration, rather than at near depletion. This regime of co-utilisation occurred with slowed xylose uptake and relatively minimal growth over the protracted 55.5 h period between $T_{90}$ and $t_{X,max}$.

For fructose, ethanol and pyruvate accumulation was minimal and delayed, with peak concentrations an order of magnitude lower than those observed for glucose and sucrose. The onset of ethanol and pyruvate production occurred between 121-145 h, coinciding with the Gompertz predicted start of the second growth phase. In contrast, trace acetate was detected persistently after $T_{10}$ and reached its peak concentration at 121 h. Fructose was also the only case in which acetate was the dominant byproduct. Byproduct reuptake occurred slowly between $T_{90}$ to $t_{X,max}$, and were co-consumed with the final 27% remaining fructose, mirroring the xylose condition.

Byproduct formation by the galactose culture was relatively low and delayed, demonstrating a similar dynamic trend to that of fructose. As such, the onset of byproducts occurred at 138 h, close to $\lambda_2$ predicted by the Gompertz model, and their period of formation coincided with that of the highest galactose uptake rate. In contrast, a higher ethanol yield was achieved compared to fructose, and acetate did not precede the other byproducts. Byproduct reassimilation occurred after $T_{90}$ over a similar timeframe to that observed for the xylose condition, with acetate being the only remaining byproduct at 254 h.

Finally, the lactose culture did not produce any detectable byproducts, despite containing a glucose moiety, which was shown to support high byproduct formation rates. This is reflected in the improved biomass yield efficiency for lactose.

### 3.1.4 Summary of single-substrate phenotypes

Comparison of the Gompertz model fits across single substrates demonstrate distinct growth dynamics. Glucose and sucrose supported rapid early growth with high $\mu_m$ with intermediate final biomass, while xylose supported slower, balanced growth over a single phase. Galactose and lactose dynamics illustrate that slow growth does not necessarily translate to high biomass accumulation as in the case of fructose and xylose and therefore may be an indicator of distinct limitations specific to lactose and its constituent monomer. Across the substrates for which a two-phase Gompertz model was selected (glucose, fructose, and galactose), an interesting similarity emerged in the timing of the transition between the first and second phases. In all three cases, the onset of the second, slower growth period occurred at comparable times ($138 < \lambda_2 < 141$). Despite this similarity in timing, the relative contribution of late-stage growth differed between substrates, which may suggest different functional explanations for the occurrence of a secondary growth phase.

Generally, an inverse relationship between substrate uptake rate and apparent biomass yield was observed across the tested carbon sources. Rapidly consumed substrates such as glucose and sucrose showed high uptake rates and early depletion, but yielded lower biomass per gram sugar, whereas slowly utilised substrates such as fructose and xylose supported extended growth periods with higher apparent yields. The galactose case somewhat deviated from this trend, demonstrating low biomass yield despite slow utilisation. Lactose was utilised at an incredibly low rate, and only partial conversion was achieved despite the lengthy experimental timeframe but did achieve a relatively high yield. These results further highlight that substrate identity strongly governs both growth dynamics and the efficiency with which assimilated carbon is converted into biomass by *F venenatum* under the studied conditions.

Byproduct kinetics in *F. venenatum* is strongly influenced by substrate type and is closely linked to substrate uptake and growth dynamics. Rapidly metabolised substrates, including glucose and sucrose, result in significant byproduct formation primarily through ethanol and pyruvate export. In contrast, slowly metabolised substrates such as fructose and galactose produced low byproduct quantities, and their onset was delayed, coinciding with the model-predicted transition to a slower secondary growth regime. Xylose represents an intermediate case, with mid-onset, moderate byproduct formation and an early transient xylitol accumulation. These distinct byproduct patterns help to provide a metabolic explanation for observed differences in biomass yield and growth dynamics among substrates, highlighting the value of analytical quantification of extracellular metabolites when evaluating growth performance in microscale batch cultures.

Overall, the single-substrate datasets demonstrate that apparent growth performance in microscale batch systems cannot be fully explained from growth rate or substrate uptake alone. Instead, substrate identity determines how carbon flux is partitioned between rapid growth, byproduct formation, and sustained biomass accumulation, giving rise to distinct conditional phenotypes.

## 3.2 Dual substrate kinetics

Figure 3 shows the time series experimental results observed for *F. venenatum* growth on glucose 15 g·L$^{-1}$ + xylose 5 g·L$^{-1}$ (A); glucose 10 g·L$^{-1}$ + xylose 10 g·L$^{-1}$ (B); glucose 15 g·L$^{-1}$ + galactose 5 g·L$^{-1}$ (C); and glucose 10 g·L$^{-1}$ + galactose 10 g·L$^{-1}$ (D). The biomass dynamics, substrate utilisation kinetics, and byproduct formation/reassimilation are discussed in the following subsections.

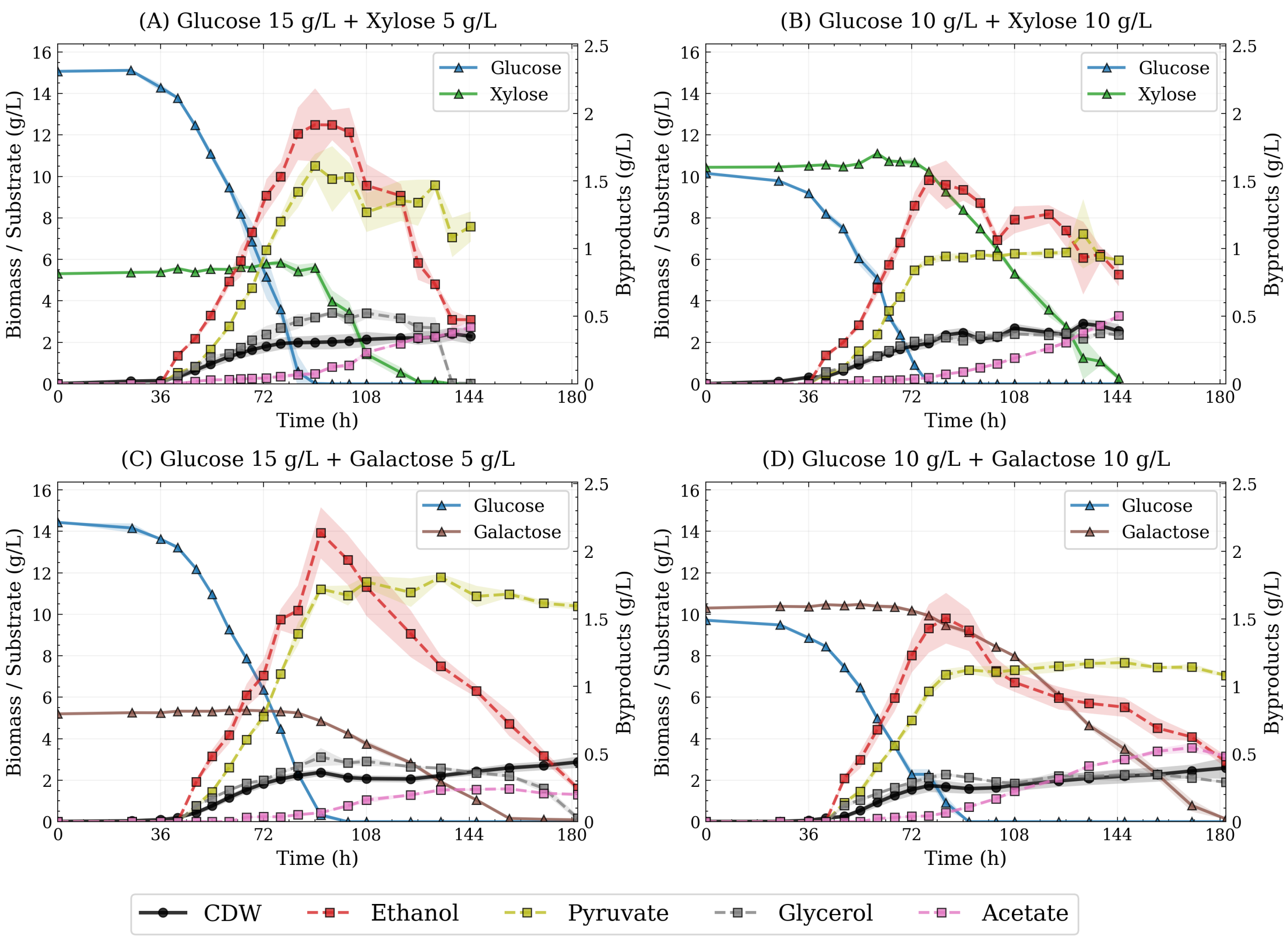

*Figure 3. Time-series experimental data of F. venenatum growth and byproduct formation / reassimilation on various synthetic dual substrates. (A) Glucose 15 g·L$^{-1}$ + Xylose 5 g·L$^{-1}$; (B) Glucose 10 g·L$^{-1}$ + Xylose 10 g·L$^{-1}$; (C) Glucose 15g·L$^{-1}$ + Galactose 5 g·L$^{-1}$; (D) Glucose 10g·L$^{-1}$ + Galactose 10 g·L$^{-1}$. Points and shaded bands represent the experimental mean and standard deviation respectively. The mean trend is indicated by a line.*

### 3.2.1 Growth dynamics across dual sugars

Across all dual-substrate conditions, Gompertz models demonstrated excellent accuracies and high goodness-of-fit across datasets ($R^2 \geq 0.98$ and RMSE < 0.16 g·L$^{-1}$), and phase number was clearly distinguished based on the ΔAICc (Table 6). The Gompertz model fits are presented in Figure 4, while key growth metric results are presented in Table 7.

*Table 6. Gompertz model selection and summary statistics for biomass growth under dual-substrate conditions. The preferred number of growth phases was selected using AICc, with ΔAICc indicating separation from the next-best model. RMSE and $R^2$ quantify fit accuracy. Glu = Glucose; Xyl = Xylose; Gal = Galactose.*

| **Condition** | **Phases** | **AICc** | **AICc (runner-up)** | **ΔAICc** | **RMSE** ($g{\cdot}L^{-1}$) | $R^2$ |
|---|---|---|---|---|---|---|
| **Glu 15 + Xyl 5g·L$^{-1}$** | 2 | -112 | -100 | 12 | 0.05 | 1 |
| **Glu 10 + Xyl 10g·L$^{-1}$** | 1 | -72 | -65 | 7 | 0.16 | 0.98 |
| **Glu 15 + Gal 5g·L$^{-1}$** | 2 | -74 | -57 | 17 | 0.11 | 0.99 |
| **Glu 10 + Gal 10g·L$^{-1}$** | 2 | -70 | -56 | 14 | 0.12 | 0.99 |

Compared with the single substrate systems, which demonstrated a wide variety of maximum biomass concentrations and dynamic growth profiles, the mixed substrate systems exhibited a trend of relatively conserved temporal trajectories, as well as similar maximum biomass concentrations, ranging from 2.41 ± 0.32 g·L$^{-1}$ to 2.9 ± 0.22 g·L$^{-1}$. Although the absolute value of the maximum biomass concentration is rather low, it is not used to compare with the previous single substrate systems due to a lower total carbon loading (20 g·L$^{-1}$ total sugar).

Maximum specific growth rates in the mixed-substrate conditions were similar to the glucose-only condition, with the glucose–galactose mixtures supporting slightly higher $\mu_m$. For example, $\mu_m$ was 0.12 h$^{-1}$ (3:1) and 0.1 h$^{-1}$ (1:1) for glucose–xylose, slightly lower than 0.13 h$^{-1}$ observed for glucose alone. On the other hand, glucose–galactose exhibited higher values of 0.14 h$^{-1}$ (3:1) and 0.13 h$^{-1}$ (1:1). It appears that for both mixtures, increasing the portion of secondary substrate resulted in a slightly lower $\mu_m$.

In glucose–xylose mixtures, maximum biomass concentrations ranged from 2.41 to 2.90 g·L$^{-1}$ and were reached between 132 and 138 h. For the 3:1 ratio, the majority of growth was contributed to the first growth phase (86%), with its transition to plateau occurring at 71 h. The estimated diauxic plateau lasted for approximately 31 h, followed by a slow increase in biomass during the second phase. For the 1:1 condition, the maximum observed biomass concentration (2.90 g·L$^{-1}$) was higher than that predicted by the model (2.47 g·L$^{-1}$), the latter of which more closely aligns with the 3:1 condition. Furthermore, this was the only condition for which the AICc favoured a single-phase description of growth. Both observations could be explained by the high temporal variability in biomass concentrations post-$T_{90}$, indicating limited resolution of a weak late-stage growth regime rather than the complete absence of secondary growth.

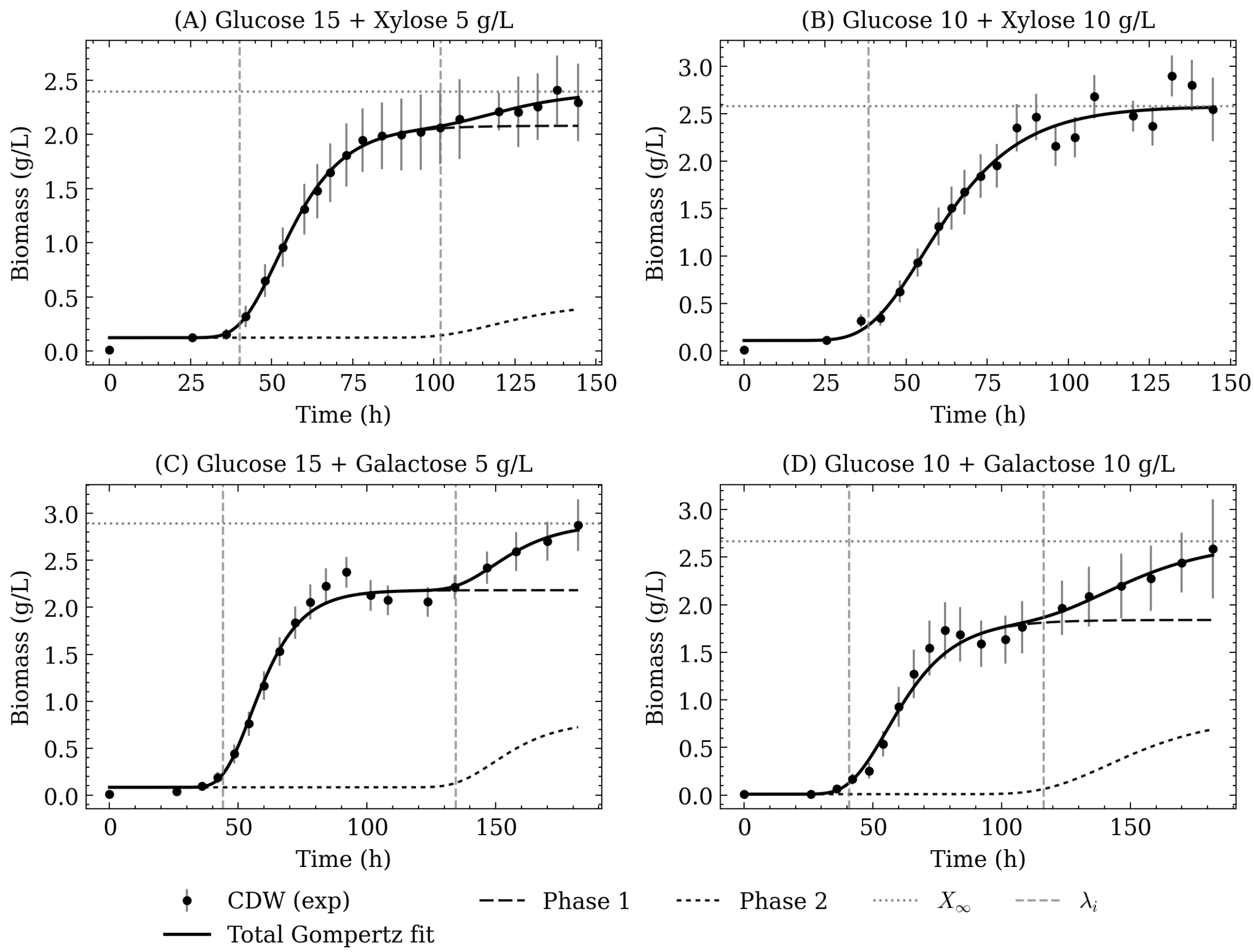

*Figure 4. Model simulated best fit modified Gompertz functions plotted against experimental CDW data for dual synthetic substrates. (A) Glucose 15 + Xylose 5g·L$^{-1}$; (B) Glucose 10 + Xylose 10g·L$^{-1}$; (C) Glucose 15 + Galactose 5g·L$^{-1}$; (D) Glucose 10 +Galactose 10g·L$^{-1}$. Black circles and vertical bars represent the mean and standard deviation of CDW respectively.*

Glucose-galactose mixtures achieved maximum biomass concentrations of 2.87 ± 0.27 g·L$^{-1}$ and 2.59 ± 0.52 g·L$^{-1}$ respectively and demonstrated much longer growth durations than the glucose-xylose conditions ($T_{90}$ ~ 157-158 h). This is a result of longer diauxic lag durations observed for these conditions ($DP$ ~70 h and 36 h for 3:1 and 1:1 respectively) and the larger contribution of the slower second phases to the overall growth (25% and 31% for 3:1 and 1:1 respectively). Although not explicitly captured by the multi-Gompertz model, there was a decline in the measured biomass concentration indicating a degree of cell death occurring during the diauxic transition, similar to the case of glucose alone, while this was not observed for the glucose-xylose conditions. It is notable that the time of the plateau transitions occur at similar times within ratios studied, regardless of the secondary substrate added ($\lambda_{DP}$ between 71 - 74 h and 80 - 84 h for 3:1 and 1:1 respectively). This indicates mostly conserved primary growth associated with glucose utilisation, with the secondary/late phase dynamics being the main differentiator between conditions.

*Table 7. Summary of growth performance metrics for dual substrates. Experimental values are reported as mean ± standard deviation. The first number in the ratio corresponds to the relative fraction of glucose in the mixture.*

| | **Glucose + Xylose (20 g·L$^{-1}$)** | | **Glucose + Galactose (20 g·L$^{-1}$)** | |
|---|---|---|---|---|
| **Metric** | **3:1** | **1:1** | **3:1** | **1:1** |
| $\boldsymbol{X_{max}}$ (g·L$^{-1}$) | 2.41±0.32 | 2.90±0.22 | 2.87±0.27 | 2.59±0.52 |
| $\boldsymbol{t_{X_{max}}}$ (h) | 138 | 132 | 182 | 182 |
| $\boldsymbol{\mu_m}$ (h$^{-1}$) | 0.12 | 0.1 | 0.14 | 0.13 |
| $\boldsymbol{\lambda_1}$ (h) | 40.1 | 38.5 | 44 | 40.1 |
| $\boldsymbol{\lambda_2}$ (h) | 102 | - | 134.5 | 116.3 |
| $\boldsymbol{\lambda_{DP}}$ (h) | 71.3 | - | 73.5 | 80.5 |
| $\boldsymbol{DP}$ (h) | 30.7 | - | 70 | 35.8 |
| $\boldsymbol{T_{10}}$ (h) | 42.9 | 42.1 | 47.2 | 45.7 |
| $\boldsymbol{T_{50}}$ (h) | 58.8 | 64.5 | 64.6 | 70.7 |
| $\boldsymbol{T_{90}}$ (h) | 117.1 | 92.7 | 157.9 | 157.4 |

### 3.2.2 Substrate utilisation kinetics

Substrate utilisation kinetics in mixtures demonstrated a strict sequential utilisation pattern, with glucose ($S_1$) preferentially consumed prior to uptake of xylose/galactose ($S_2$), consistent with diauxic growth. Across all conditions, glucose depletion occurred rapidly and early, after which utilisation of the secondary carbon source proceeded at substantially lower rates with reduced apparent efficiencies. Key substrate utilisation metrics are presented in Table 8.

In glucose–xylose mixtures, glucose depletion occurred between 78 and 90 h, with faster depletion observed at the 3:1 ratio relative to the 1:1 condition. The relative depletion times align well with the $\lambda_{DP}$values estimated from the Gompertz models. Xylose utilisation commenced immediately after the depletion of glucose in both cases, proceeding at near-identical, slower uptake rates (~0.15 g·L$^{-1}$·h$^{-1}$).

Glucose–galactose mixtures displayed analogous sequential utilisation behaviour, yet with distinct secondary substrate uptake kinetics. Glucose depletion occurred between 84 and 92 h, again in general agreement with the Gompertz predicted onset times of the diauxic plateau. Galactose utilisation began subsequent to glucose depletion, but at lower rates when compared with xylose (0.061 and 0.1 g·L$^{-1}$·h$^{-1}$ for the 3:1 and 1:1 conditions respectively), reaching depletion between 170 and 182 h.

*Table 8. Summary of substrate utilisation kinetics for duals substrates. Experimental values are reported as mean ± standard deviation. * Xylose not fully depleted within the experimental timeframe.*

| | **Glucose + Xylose (20g·L$^{-1}$)** | | **Glucose + Galactose (20g·L$^{-1}$)** | |
|---|---|---|---|---|
| **Metric** | **3:1** | **1:1** | **3:1** | **1:1** |
| $\boldsymbol{S_{1,in}}$ (g·L$^{-1}$) | 15.07±0.05 | 10.16±0.08 | 14.42±0.04 | 9.77±0.06 |
| $\boldsymbol{t_{S1\sim 0}}$ (h) | 90 | 78 | 92 | 84 |
| $\boldsymbol{qS_{1,avg}}$ (g·L$^{-1}$·h$^{-1}$) | 0.285 | 0.224 | 0.238 | 0.165 |
| $\boldsymbol{Y_{X/S1}}$ (g·g$^{-1}$) | 0.14 | 0.2 | 0.17 | 0.16 |
| $\boldsymbol{S_{2,in}}$ (g·L$^{-1}$) | 5.31±0.01 | 10.43±0.04 | 5.23±0.03 | 10.29±0.03 |
| $\boldsymbol{t_{S2\sim 0}}$ (h) | 132 | -* | 170 | 182 |
| $\boldsymbol{qS_{2,avg}}$ (g·L$^{-1}$·h$^{-1}$) | 0.152 | 0.151 | 0.061 | 0.1 |
| $\boldsymbol{Y_{X/S2}}$ (g·g$^{-1}$) | 0.04 | 0.06 | 0.07 | 0.09 |
| $\boldsymbol{Y_{X/S}}$ (g·g$^{-1}$) | 0.12 | 0.13 | 0.14 | 0.13 |

The faster uptake rate of glucose observed for the 3:1 conditions suggest a proportional relationship between glucose loading and rate of uptake. Indeed, the glucose-only condition studied at 30g·L$^{-1}$ demonstrated an even higher rate of uptake (0.538 g·L$^{-1}$·h$^{-1}$). This is likely a simple result of higher concentration gradient as a driving force for more rapid cross-membrane transport.

Across all conditions, the overall apparent biomass yields were similarly low, ranging from 0.12 – 0.14 g·g$^{-1}$. In terms of relative yield efficiencies, biomass was most efficiently formed from glucose (0.14 – 0.2 g·g$^{-1}$), while the yield from secondary substrates was lower, with galactose efficiencies being higher than that of xylose.

Critically, the low yield efficiencies associated with the secondary substrates contrast with the apparent biomass yields determined for their corresponding single-substrate cultures. When supplied as the only available sugar, xylose and galactose achieved much higher biomass yields than when supplied in mixture with glucose. This is due to the observed expenditure of secondary carbon during the diauxic transition period that does not contribute to biomass formation. For both mixtures, the yield efficiency of the secondary substrate increased when supplied at the same concentration as glucose. The difference in the diauxic lag adaptation times between ratios of glucose-galactose support this discrepancy. The plateau duration for the 3:1 condition was estimated to be ~70 h, approximately double the time estimated for the 1:1 condition. This reflects a longer period of poor anabolic efficiency for the secondary substrate when supplied at a lower initial concentration.

### 3.2.3 Byproduct formation and reassimilation

Key metrics for byproduct formation and reassimilation in dual substrates are presented in Table 9 with emphasis on ethanol as the predominant byproduct. Comprehensive metrics are presented in Supplementary Information 4.

In all dual-substrate conditions, ethanol, pyruvate and glycerol accumulation was tightly associated with the period of glucose utilisation during the primary phase, with $t_{EtOH_{max}}$ coinciding precisely with $t_{S1\sim0}$ and close to $\lambda_{DP}$. Similar peak values and formation rates were also observed across substrate combinations and ratios, while ethanol formation yields varied modestly between conditions (0.12 to 0.18 $g \cdot g^{-1}$). Generally, ethanol was the predominant byproduct in all cases, followed by pyruvate and glycerol, whereas in both 1:1 conditions, the maximum acetate concentration exceeded that of glycerol.

Interestingly, it appears that the rate and degree of byproduct reassimilation was influenced by the initial concentration of the secondary substrate in both mixtures. Indeed, both 3:1 mixtures demonstrated higher byproduct reuptake rates across the board, and more complete byproduct reassimilation within the experimental timeframe. Glycerol provides a clear example of this differentially repressed reuptake by the secondary substrate. In both 3:1 ratios, glycerol declined only marginally during consumption of xylose/galactose. However, the uptake rate increased around the time of total sugar depletion, and 100% reassimilation was achieved. It can therefore be inferred that the onset, rate and extent of byproduct reoxidation is a function of secondary sugar presence and concentration and varies with byproduct identity, with pyruvate and glycerol uptake supressed to a high degree in the presence of the secondary sugar, and ethanol uptake repressed the least.

Table 9. Byproduct formation and reassimilation metrics for dual substrates. Experimental values are reported as mean ± standard deviation. EtOH = ethanol; Pyr = pyruvate; Gly = glycerol; Ace = acetate.

| | Glucose + Xylose (20g·L$^{-1}$) | | Glucose + Galactose (20g·L$^{-1}$) | |
|---|---|---|---|---|
| **Metric** | **3:1** | **1:1** | **3:1** | **1:1** |
| $EtOH_{max}$ (g·L$^{-1}$) | 1.92±0.27 | 1.51±0.09 | 2.14±0.19 | 1.51±0.19 |
| $t_{EtOH_{max}}$ (h) | 90 | 78 | 92 | 84 |
| $qP_{EtOH,max}$ (g·L$^{-1}$·h$^{-1}$) | 0.054 | 0.049 | 0.047 | 0.046 |
| $Y_{EtOH/S}$ (g·g$^{-1}$) | 0.12 | 0.16 | 0.16 | 0.18 |
| $qP_{reassim,EtOH}$ (g·L$^{-1}$·h$^{-1}$) | 0.03 | 0.011 | 0.021 | 0.011 |
| $Pyr_{max}$ (g·L$^{-1}$) | 1.61±0.09 | 1.11±0.26 | 1.81±0.03 | 1.18±0.05 |
| $Gly_{max}$ (g·L$^{-1}$) | 0.53±0.01 | 0.37±0.02 | 0.48±0.07 | 0.35±0.01 |
| $Ace_{max}$ (g·L$^{-1}$) | 0.42±0.03 | 0.5±0.01 | 0.24±0.01 | 0.55±0.03 |

A distinctive and consistent feature of the dual-substrate datasets was the delayed onset of acetate accumulation. In all conditions, acetate remained at low or near-background levels during the primary glucose utilisation phase and began to increase only after glucose depletion, coinciding with the onset of ethanol reassimilation and secondary substrate uptake. Acetate concentrations continued to rise throughout the secondary phase and exhibited no obvious reassimilation phase within the experimental times.

### 3.2.4 Summary of dual-substrate phenotypes

Overall, the dual-substrate growth phenotype was characterised by rapid glucose-supported biomass accumulation followed by a slower secondary growth phase. While both xylose and galactose supported secondary growth following glucose depletion, galactose-containing mixtures exhibited a more pronounced and extended transition between growth regimes, including a detectable period of biomass loss prior to secondary growth. These distinctions demonstrate that secondary substrate identity strongly influences not only the extent but also the structure and continuity of late-stage growth under mixed-carbon conditions, even when total biomass accumulation remains similar.

Glucose depletion occurred between 84 and 92 h across both ratios, with uptake rates decreasing as glucose fraction was reduced. Galactose uptake was markedly slower than xylose uptake under equivalent ratios, with substantially protracted times to depletion.

Apparent biomass yields associated with galactose utilisation were slightly higher than those for xylose under mixed-substrate conditions but remained low overall. Across both substrate pairs, increasing the relative amount of the secondary substrate modestly increased the apparent efficiency of its utilisation, but this did not translate into a proportional increase in overall biomass yield. As a result, global biomass yields for glucose–galactose mixtures were comparable to those observed for glucose–xylose mixtures and showed limited sensitivity to substrate ratio. These results indicate that although secondary substrates were taken up following glucose depletion with little apparent delay, their utilisation occurred at reduced rates and low anabolic efficiency relative to both glucose and their respective single-substrate counterparts.

Byproduct profiles demonstrated a relatively conserved metabolic sequence across substrate combinations. That being rapid glucose-associated fermentative production dominated by ethanol and pyruvate with a small glycerol contribution. This was followed by a post-glucose phase in which ethanol, pyruvate and glycerol are either partially reassimilated or remain at constant levels over the remainder of the time series, with the degree of reassimilation and uptake repression varying with the concentration of the secondary substrate present and the type of byproduct. Therefore, byproduct reassimilation dynamics are shaped not only by prior glucose metabolism but also by the concurrent presence and utilisation of the secondary sugar.

## 3.3 Expired functional drink and synthetic control comparison

Figure 5 shows the time series experimental results observed for *F. venenatum* growth on expired functional drink (A) and its synthetic control (B) formulated to match the measured sugar and citric acid composition of the real sample. The biomass dynamics, substrate utilisation kinetics, and byproduct formation/reassimilation are discussed in the following subsections.

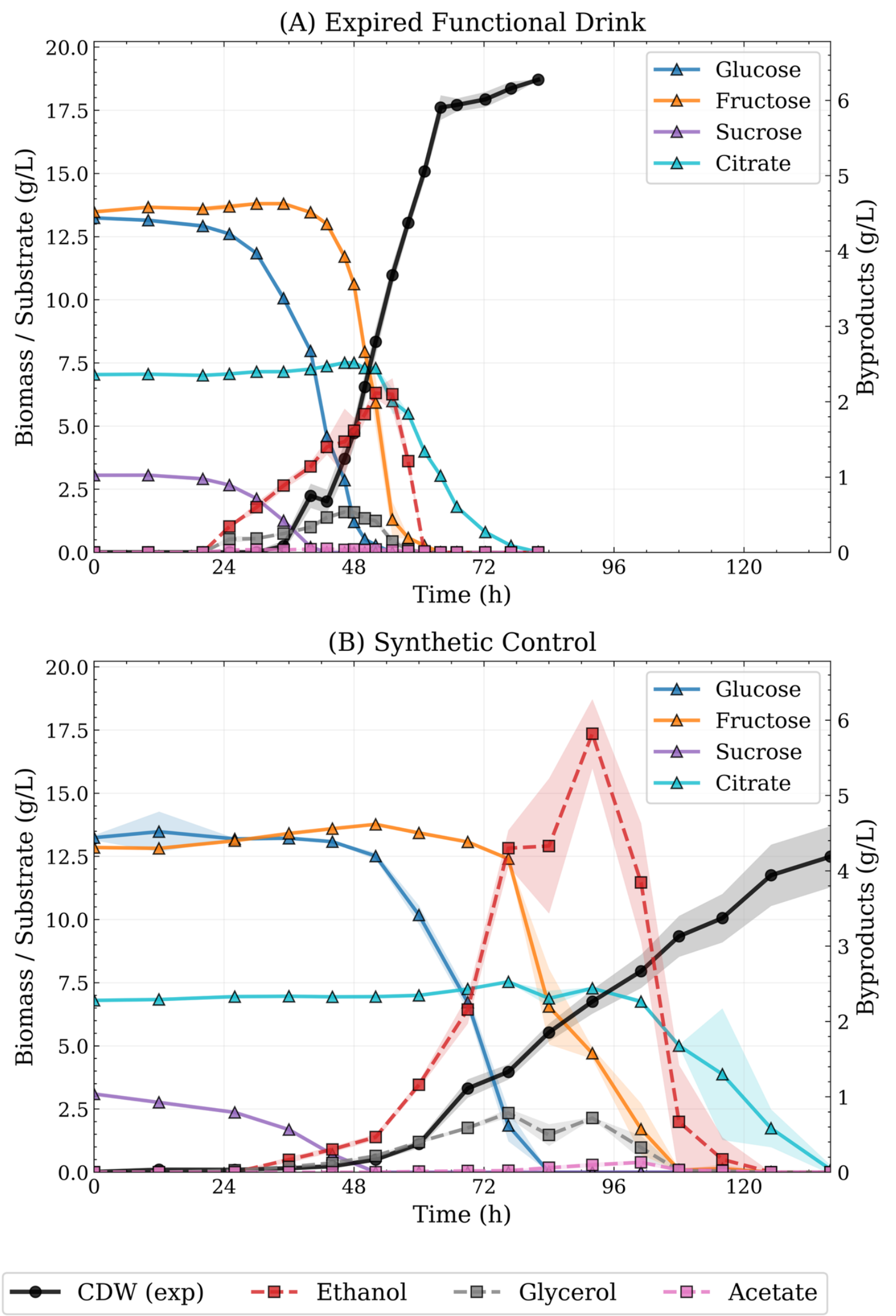

*Figure 5. Time-series experimental profiles of F. venenatum growth and byproduct formation / reassimilation from expired functional drink and a synthetic control. (A) Expired functional drink; (B) Synthetic control. Points and shaded bands represent the experimental mean and standard deviation respectively. The mean trend is indicated by a line.*

### 3.3.1 Growth dynamics for expired functional drink and synthetic control

The growth dynamics for both the real sample and control condition were best described by a single phase Gompertz model, favoured by the AICc to similar extents when compared with two phase fits. Both estimated models demonstrated high goodness of fit, with $R^2$ near unity (0.99) and low RMSE values for both conditions. Summary statistics for these conditions are presented in Table 10, model fits are presented in Figure 6 and Gompertz derived parameters are presented in Table 11.

*Table 10. Gompertz model selection and summary statistics for biomass growth on expired functional drink and synthetic control. The preferred number of growth phases was selected using AICc, with ΔAICc indicating separation from the next-best model. RMSE and $R^2$ indicate goodness of fit.*

| Condition | Phases | AICc | AICc (runner-up) | ΔAICc | RMSE ($g \cdot L^{-1}$) | $R^2$ |
|---|---|---|---|---|---|---|
| **Expired Functional Drink** | 1 | -22 | -12 | 10 | 0.64 | 0.99 |
| **Synthetic Control** | 1 | -26 | -14 | 12 | 0.38 | 0.99 |

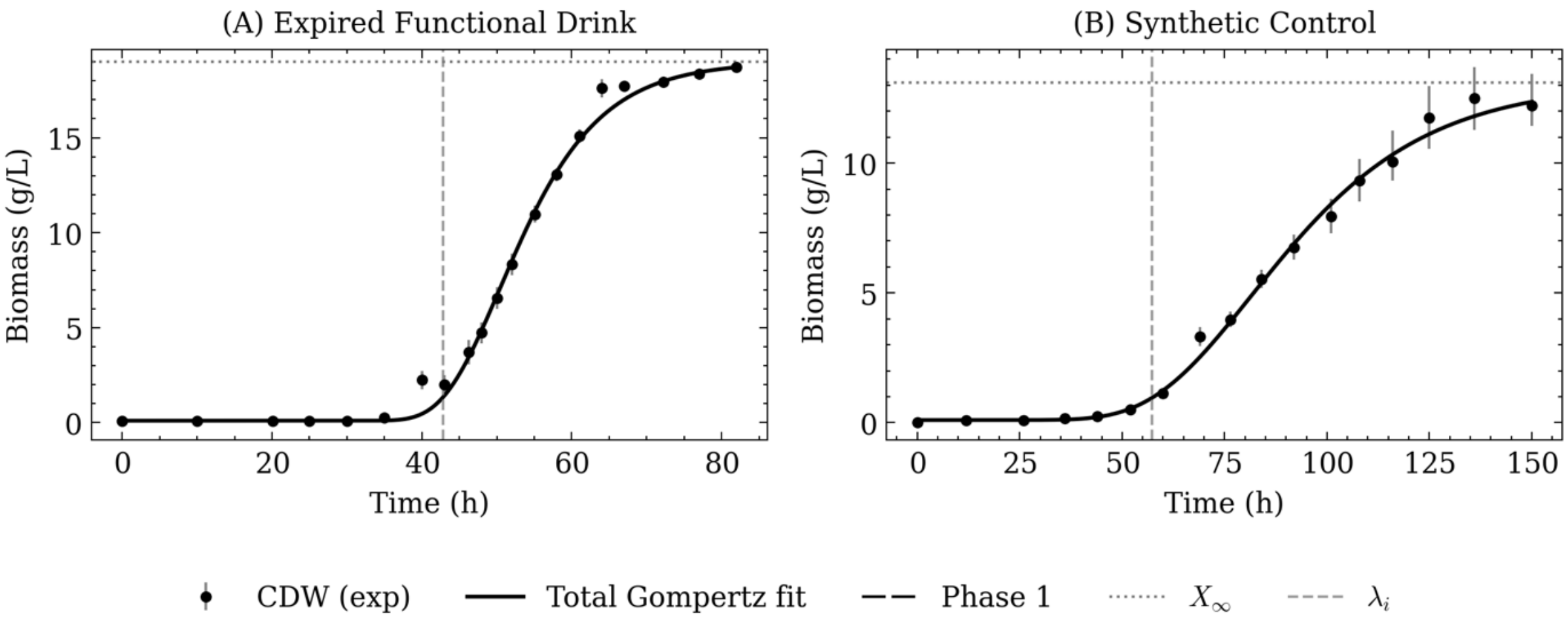

*Figure 6. Model simulated best fit modified Gompertz functions plotted against experimental CDW data for expired functional drink and its synthetic control. (A) Expired functional drink; (B) Synthetic control. Black circles and vertical bars represent the mean and uncertainty of the CDW respectively.*

The real expired functional drink sample supported an exceptionally high maximum biomass titre, reaching 18.37 ± 0.64 $g \cdot L^{-1}$ after 82 h. The synthetic control condition achieved a lower maximum biomass of 12.49 ± 1.21 $g \cdot L^{-1}$ over a longer period (136 h). In addition, after a relatively short lag phase similar to that found for synthetic glucose at 30 $g \cdot L^{-1}$ (42.8 h), the real sample demonstrated a rapid growth pace from the onset through $T_{90}$.

Compared with the synthetic control and all single and dual substrate conditions studied, the real sample exhibited the highest estimated $\mu_m$ (0.33 $h^{-1}$) and the fastest progression through growth milestones, with $T_{90}$ reached ~23 h following the lag phase. In contrast, the synthetic condition supported an approximately 3-fold lower maximum specific growth rate following a longer lag period (57.3 h) and exhibited slower overall growth, requiring nearly twice the time to reach $T_{90}$ (126.8 h).

While the expired functional drink sample achieved more rapid growth and higher biomass titre compared with the synthetic control, consideration of the corresponding substrate uptake profiles reveals that biomass accumulation and substrate depletion proceed asynchronously during the late stage of cultivation. Interestingly in both cases, the lack of identifiability of multiple phases from Gompertz analysis despite mixed substrate complexity highlights the need for resolved quantification of the metabolite concentrations over time in order to understand carbon allocation strategies leading to this observed growth profile.

The following subsection therefore focuses on substrate utilisation dynamics to contextualise observed growth behaviour and to assess the robustness of the apparent biomass yields.

*Table 11. Summary of growth performance metrics for expired functional drink and synthetic control. Experimental values are reported as mean ± standard deviation.*

| Metric | Expired Functional Drink | Synthetic Control |
| --- | --- | --- |
| $\boldsymbol{X_{max}}$ (g·L$^{-1}$) | 18.37 ± 0.64 | 12.49 ± 1.21 |
| $\boldsymbol{t_{X_{max}}}$ (h) | 82 | 136 |
| $\boldsymbol{\mu_m}$ (h$^{-1}$) | 0.33 | 0.1 |
| $\boldsymbol{\lambda}$ (h) | 42.8 | 57.3 |
| $\boldsymbol{T_{10}}$ (h) | 44 | 60 |
| $\boldsymbol{T_{50}}$ (h) | 53 | 88.8 |
| $\boldsymbol{T_{90}}$ (h) | 66.4 | 126.8 |

### 3.3.2 Substrate utilisation kinetics

The compositional characterisation of the expired functional drink sample found similar concentrations of glucose (13.24 ± 0.03 g·L$^{-1}$) and fructose (13.48 ± 0.06 g·L$^{-1}$), and a smaller fraction of total sugar contributed by sucrose (3.06 ± 0.01 g·L$^{-1}$). Citric acid was present at a concentration of 7.04 ± 0.01 g·L$^{-1}$. Overall, this sample had a high carbon load and was compositionally complex. Table 12 summarises the compositional characterisation of the expired functional drink sample, as well as the substrate utilisation metrics compared with the synthetic control.

*Table 12. Summary of substrate utilisation kinetics for expired functional drink and synthetic control. Experimental values are reported as mean ± standard deviation.*

| Metric | Expired Functional Drink | Synthetic Control |
|---|---|---|
| $S_{Glu,in}$ ($g \cdot L^{-1}$) | 13.24 ± 0.03 | 13.24 ± 0.1 |
| $t_{Glu \approx 0}$ (h) | 52 – 55 | 76.5 – 84 |
| $qS_{Glu,avg}$ ($g \cdot L^{-1} \cdot h^{-1}$) | 0.681 | 0.435 |
| $S_{Fru,in}$ ($g \cdot L^{-1}$) | 13.48 ± 0.06 | 12.85 ± 0.05 |
| $t_{Fru \approx 0}$ (h) | 61 – 64 | 101 – 108 |
| $qS_{Fru,avg}$ ($g \cdot L^{-1} \cdot h^{-1}$) | 1.178 | 0.437 |
| $S_{Suc,in}$ ($g \cdot L^{-1}$) | 3.06 ± 0.01 | 3.1 ± 0.03 |
| $t_{Suc \approx 0}$ (h) | 40 – 43 | 44 – 52 |
| $qS_{Suc,avg}$ ($g \cdot L^{-1} \cdot h^{-1}$) | 0.136 | 0.093 |
| $CA_{in}$ ($g \cdot L^{-1}$) | 7.04 ± 0.01 | 6.8 ± 0.03 |
| $t_{CA \approx 0}$ (h) | 77 – 82 | 125 – 136 |
| $qCA_{avg}$ ($g \cdot L^{-1} \cdot h^{-1}$) | 0.32 | 0.19 |
| $Y_{\frac{X}{S+CA}}$ (C-mol/C-mol) | 0.54 | 0.36 |

As for the case of synthetic sucrose at 30 $g \cdot L^{-1}$, sucrose in both the real and control samples was hydrolysed during the initial lag period. Complete hydrolysis coincided with the onset of sustained growth predicted by the Gompertz fits, and sucrose in the real sample was hydrolysed more rapidly than that of the control (0.136 $g \cdot L^{-1} \cdot h^{-1}$ vs. 0.093 $g \cdot L^{-1} \cdot h^{-1}$). In both cases the rate of sucrose hydrolysis was dramatically lower than that observed for synthetic sucrose at 30$g \cdot L^{-1}$ (~0.82 $g \cdot L^{-1} \cdot h^{-1}$), indicative of the effect of total sucrose concentration on the rate of hydrolysis by secreted enzymes (i.e. invertase).

The observed sequence of preferential uptake of glucose followed by fructose is present in both conditions (as in the sucrose 30$g \cdot L^{-1}$ condition), however, the degree of glucose repression on fructose uptake is not strict and the extent of co-utilisation differed between the two conditions. For the real sample, by 46 h the fructose concentration had declined below 85% of its peak concentration with 2.86 $g \cdot L^{-1}$ residual glucose remaining. For the synthetic control, the transition between glucose and fructose utilisation was more distinct, with fructose concentration only declining to 90% of its peak concentration with 1.85 $g \cdot L^{-1}$ residual glucose remaining. Both glucose and fructose were consumed more rapidly in the case of the real sample. Glucose uptake was ~57% higher than in the synthetic control, with the rate of fructose uptake showing the most dramatic increase at ~170% the rate of the synthetic control. Fructose uptake in the real sample also demonstrated the highest rate of all conditions studied thus far, outpacing synthetic fructose at 30 $g \cdot L^{-1}$ by nearly 10-fold. These observations indicate the real sample not only supports faster sugar utilisation rates but also reduced time-separation between hexose utilisation phases, which may in turn reflect lower catabolite repression effects of glucose on fructose uptake in the real system.

Interestingly, citric acid appears to act as a fourth carbon source in both systems and is consumed in the latter stages of the cultivation post-sugar depletion. Initial limited uptake aligned with fructose declining below a threshold concentration and the early period of ethanol reoxidation. There was also a discrepancy in the rate of citric acid uptake, with the real sample supporting a near two-fold higher rate at 0.32 $g \cdot L^{-1} \cdot h^{-1}$.

Despite the higher rates of uptake for each of the four carbon sources, the real sample was able to achieve a higher apparent biomass yield compared to the control (0.54 vs. 0.36 C-mol/C-mol) from total supplied sugars and citric acid. This finding is in contrast with previously observed trends for synthetic media conditions which generally found that higher substrate uptake rates were associated with lower biomass yields. However, while this system-level metric reflects the overall conversion of supplied carbon to final CDW, it does not resolve how the biomass was formed over time or attributable to individual substrates.

This leads to the final critical observation for these conditions regarding the relationship between the apparent biomass yield and the temporal sequence of carbon utilisation. The growth window coinciding approximately with the period between $T_{10}$ and $T_{50}$ is associated primarily with utilisation of hexose carbon through glucose and fructose uptake, with only minor initial reoxidation of glycerol. During this phase, the apparent biomass yield is 0.58 C-mol/C-mol (0.56 $g \cdot g^{-1}$), indicating a period of highly efficient biomass formation from these sugars within realistic stoichiometric bounds [21]. This is followed by a transitional period between 52 and 61 h where carbon uptake cannot be cleanly allocated to a single carbon source due to overlapping utilisation of residual fructose, ethanol, and citric acid. In contrast, the final substrate-defined growth period from 61 – 84 h was characterised by continued and substantial biomass accumulation concurrent with citric acid uptake as the only remaining measured external source of carbon. In isolation, this window would imply an apparent biomass yield of ~0.9 C-mol/C-mol which is physiologically implausible. Indeed, the efficiency of cellular biosynthesis on citric acid is expected to be lower than that of sugars due to its lower degree of reduction.

This late stage increase in the measured biomass cannot therefore be attributed to citric acid utilisation and/or primary cellular growth alone and reflects a decoupling of CDW increase and the measured substrate uptake. This likely suggests uncaptured contributions of unquantified carbon source(s) or changes in biomass composition during late-stage growth. Importantly, this behaviour would not be evident from end-point measurements alone. By resolving both substrate and biomass time series data, it is possible to identify periods in which apparent yield estimates are most robust and plausible and those where they overstate true biosynthetic efficiencies from the data at hand. This distinction is critical for interpreting performance metrics in compositionally complex feedstocks and is explored further in the discussion.

#### 3.3.3 Byproduct formation and reassimilation

Interesting trends emerged in the byproduct formation and reassimilation kinetics that may help to explain the differences between the two studied conditions, while contrasting with observed findings for well-defined synthetic substrates. Metrics are presented in Table 13.

Across both conditions, fermentative byproducts were produced to different extents. Ethanol was the predominant byproduct with early onset co-production with glycerol occurring after 24 h of incubation. Acetate remained at very low concentrations in both cases, and pyruvate was notably absent. The production of ethanol and glycerol coincided primarily with the utilisation of sucrose and glucose. However, in the case of the synthetic control, 26% of the maximum ethanol concentration was formed during a period of fructose-only utilisation (76.5 - 92 h), indicative of a partial fructose contribution to the overall ethanol yield.

The maximum byproduct concentrations were significantly lower for the real sample, with ethanol reaching a titre of 2.12 ± 0.07 $g \cdot L^{-1}$ compared with 5.82 ± 0.46 $g \cdot L^{-1}$ in the synthetic control. This is reflected in the hexose-equivalent ethanol yields calculated over the period of formation, with $Y_{EtOH/S}$ = 0.08 $g \cdot g^{-1}$ and 0.24 $g \cdot g^{-1}$ for the real sample and control condition respectively. The maximum rate of ethanol production achieved was also higher in the case of the synthetic control (0.19 $g \cdot L^{-1} \cdot h^{-1}$), indicating a more intense product formation period. Reuptake of ethanol following $t_{EtOH_{max}}$ was ~51% faster in the case of the real sample, indicating a higher metabolic capacity for reoxidation of fermentative byproducts in this system. Overall, these patterns suggest a more respiratory phenotype and/or improved redox balancing in the real expired functional drink condition compared with the synthetic control.

*Table 13. Byproduct formation and reassimilation metrics for expired functional drink and synthetic control. Experimental values are reported as mean ± standard deviation. EtOH = ethanol; Gly = glycerol; Ace = acetate.*

| Metric | Expired Functional Drink | Synthetic Control |
|---|---|---|
| $\boldsymbol{EtOH_{max}}$ ($g\cdot L^{-1}$) | 2.12 ± 0.07 | 5.82 ± 0.46 |
| $\boldsymbol{t_{EtOH_{max}}}$ (h) | 52 | 92 |
| $\boldsymbol{qP_{EtOH,max}}$ ($g\cdot L^{-1}\cdot h^{-1}$) | 0.125 | 0.19 |
| $\boldsymbol{Y_{\frac{EtOH}{S}}}$ ($g\cdot g^{-1}$) | 0.08 | 0.24 |
| $\boldsymbol{qP_{reassim,EtOH}}$ ($g\cdot L^{-1}\cdot h^{-1}$) | 0.355 | 0.235 |
| $\boldsymbol{Gly_{max}}$ ($g\cdot L^{-1}$) | 0.54 ± 0.01 | 0.79 ± 0.06 |
| $\boldsymbol{Ace_{max}}$ ($g\cdot L^{-1}$) | < 0.1 | 0.13 ± 0.03 |

### 3.3.4 Summary of expired functional drink phenotype

Overall, the expired functional drink supported rapid growth, high biomass titre, and efficient carbon conversion relative to its compositionally matched synthetic control. With regards to substrate utilisation, the expired drink exhibited faster uptake of all available carbon sources in addition to increased degree of co-utilisation of carbon sources. The apparent yield to biomass was higher on a hexose-only and total carbon molar basis, and there was a reduction in fermentative byproduct formation and enhanced ethanol reassimilation. These findings demonstrate that expired functional drink may function as a highly effective feedstock for microbial protein production from the perspective of productivity and yield.

However, time-series state analysis revealed behaviour that cannot be explained by the measured substrate composition alone using the current analytical method. Analysis of late-stage biomass accumulation arising from citric acid uptake led to apparent yields exceeding that which could be deemed plausible for citric acid metabolism, indicating a decoupling between measured external carbon utilisation and observed biomass increase.

These findings highlight both the potential of compositionally complex waste feedstocks for *F. venenatum* to produce mycoprotein rapidly and efficiently, as well as the limitations of end-point yield and curve-only metrics for interpreting overall performance.

# 4 Discussion

## 4.1 Rate-dependent overflow as a system-level constraint

One of the most consistent trends across all datasets was that early-onset, intense exponential growth dynamics and high substrate uptake rates were associated with increased production of fermentation byproducts (ethanol, glycerol, pyruvate and acetate) by *F. venenatum*, and a correspondingly low biomass yield.

This behaviour closely resembles that of 'overflow metabolism', a long-documented respiro-fermentative strategy exhibited by many fungal and yeast species, with *Saccharomyces cerevisiae* being the most notable example [24, 25]. This strategy occurs when substrate uptake rates are sufficiently high to exceed respiratory capacity, leading to an accumulation of cytosolic NADH from glycolysis. In order to maintain high glycolytic flux and continued energy generation, NAD+ is rapidly regenerated through fermentative conversion of pyruvate to ethanol. This leads to high growth rates but at the expense of immediate biomass yield, with overflow byproducts acting as a surplus extracellular energy source. Importantly, this mechanism has also been proven in the aerobic filamentous fungus *Aspergillus oryzae* and *Fusarium lini* [26-28], lending further support to the presence of overflow metabolism in fungi such as F. venenatum.

While fermentative byproducts were measured for all conditions to some degree (besides lactose), glucose-supported growth consistently produced the highest quantities, with ethanol being most predominant. The influence of carbon source on the extent of fermentation has been of great research interest, particularly in strains where the target molecule is ethanol in the production of biofuels. With regards to the relative influence of carbon sources, our results align well with those reported for white rot fungus *Lenzites betulinus*, which showed highest ethanol yield efficiency on glucose (0.38 $g \cdot g^{-1}$) and moderate efficiency with xylose (0.26 $g \cdot g^{-1}$), while galactose demonstrated very poor yield (0.07 $g \cdot g^{-1}$) [29]. This order is also mostly conserved when comparing our findings with studies in *S. cerevisiae* [30]. However, a notable exception is fructose. While fructose supports comparable ethanol yields to glucose in *S. cerevisiae*, *F. venenatum* produced only minimal overflow byproducts titres from fructose in the late stages of growth. This is consistent with our observation of fructose being utilised at a much lower rate when compared with glucose, with more protracted growth and higher biomass yield. This may indicate a difference in carbon utilisation pathways for glucose and fructose despite molecular similarities, which will be discussed further in the discussion.

Furthermore, while xylose exhibited signatures of overflow metabolism, the phenotypes observed for xylose-supported biomass growth were relatively distinct when compared with other substrates, likely arising from xylose-specific metabolic limitations during early growth. The early onset and transient xylitol production that occurred may be a result of redox imbalance from xylose flux through the xylose-oxidoreductase pathway, a phenomenon previously reported in recombinant *S. cerevisiae*, where limited NAD+ regeneration capacity led to accumulation of xylitol [31]. Critically, the concentration of xylitol declined in the system as the xylose uptake increased. These observations may indicate a gradual upregulation and coordination of the enzymatic pathways required to facilitate high xylose uptake rates, with full capacity achieved by the mid-exponential growth phase. After this point, fermentative byproducts were produced, indicating a shift from xylose pathway-specific redox limitations to overflow metabolism.

Although the trends in our data and discrepancies between substrates suggest that byproduct formation results primarily from overflow metabolism in the above cases, it may also be the case that oxygen transfer limitations resulted in an anaerobic growth environment in the microplate system. For instance, another plausible explanation for the dynamics of the xylose condition may be explained by a shift from $O_2$ sufficiency to limitation. For example, in xylose metabolism by *A. niger*, it was found that xylitol accumulation occurred during the aerobic exponential phase, and that acids and polyols were produced later during the oxygen limitation stage of the fermentation [32]. Additionally, the consistent co-export of glycerol in ethanol producing conditions could be linked to the extent of fermentative contribution, as it is associated with redox balance and cell osmolarity maintenance during alcoholic fermentation[33, 34]. Therefore, analysis of the co-existence, sequence of production and transience of key byproducts may serve as an indicator of the onset and extent of anaerobic growth in *F. venenatum*.

Furthermore, a consistent observational trend was the formation of a mycelial mat at the air-liquid interface during late-stage growth in conditions where hyphal development was supported (all but lactose), which has been found in other fungal species to be indicative of a stress response to hypoxic conditions [35]. However, this observation does not explain the early onset of byproduct formation by glucose and sucrose in particular, lending credence to the overflow theory for these substrates at least in the early stages of growth.

Therefore, further experiments should be employed to investigate the potential causes of byproduct formation across conditions, as sufficient aeration is critical to the transferability of kinetic findings across scales. For instance, the use of different microplate well geometries and orbital shaking strategies may significantly increase oxygen transfer rate (OTR) to the culture media, while use of breathable membranes in place of plate lids can improve oxygen transfer to the plate headspace [36]. Although not formally presented here, during protocol optimisation, the use of a polyurethane membrane (Breathe-Easy®, Sigma Aldrich) designed for aerobic cultures was employed as a plate covering and compared with glucose 30 $g \cdot L^{-1}$ using the current parafilm-sealed plate lid method. However, our results revealed no improvement in the biomass yield and in fact, a significant increase in ethanol production was observed. These preliminary findings presented in Supplementary Information 5 indicate that the extent of byproduct formation was not governed entirely by a decrease in dissolved oxygen (DO). However, quantification of the time-varying DO and estimation of the oxygen transfer coefficient (*kLa*) in culture wells should be quantified to confirm this, for example through use of optical measurements of oxygen-reactive fluorophores [37].

## 4.2 Environmental limitations and late-stage carbon reallocation may explain non-diauxic growth regime shifts

Byproduct formation by *F. venenatum* on fructose and galactose deviated from the other conditions, in that they only appeared transiently and at low concentrations during late-stage growth, coinciding with the modelled shift to a second, slow and protracted growth phase in both cases. Furthermore, these were the only two conditions wherein the Gompertz predicted second phase did not coincide with a long diauxic plateau or shift to utilisation of another carbon source. Both fructose and galactose were utilised at relatively low rates and demonstrated low maximum specific growth rates. While it is possible that the underlying cause of this byproduct formation is overflow metabolism, this would generally only be expected for cultures with higher substrate uptake and growth rates. Therefore, a plausible explanation could be that that they were produced in response to unmeasured environmental stresses, such as oxygen, nitrogen or other nutrient limitations. For instance, the oxygen transfer rate may become limiting in prolonged cultures as discussed prior, particularly as the mycelial density increases [38], while slower and yield-efficient utilisation of substrate (particularly in the case of fructose) may lead to a complete depletion of nitrogen in the system prior to total carbon utilisation, shifting the growth from carbon-limited to nitrogen-limited. For many fungal species, the term ‘idiophase’ has been used to describe a prolonged, late-stage transition to stationary phase, where under certain environmental stressors or nutrient depletion, carbon allocation shifts from synthesis of cellular biomass to that of secondary metabolites not required for growth. These may include storage compounds, lipids, polysaccharides (e.g. glycogen or extracellular polymeric substances), organic acids, alcohols, and polyketide- or terpenoid-derived compounds [39-42]. For example, in *R. mucilaginosa* and *A. oryzae*, nitrogen limitation induced elevated lipogenesis rates in the early-to-late stationary phase, resulting in structural changes in biomass composition and morphology [43, 44]. While in *F. graminaerum*, oxygen-limitation imposed by pellicle formation led to increasing lipid content of biomass over time and enhanced production of extracellular matrix compounds [45].

Therefore, the coincidence of byproducts measured at the onset of the slow second Gompertz phase in the fructose and galactose cases may indicate a period of idiophasic metabolism as a result of environmental limitations other than carbon availability.

## 4.3 Glucose and fructose are chemically similar with divergent phenotypes

Although glucose and fructose are chemically similar hexoses, they supported surprisingly distinct growth behaviours in this study. Fructose supported prolonged, slow growth, with low uptake rates and high biomass yield compared with glucose. Differences in their uptake kinetics and carbon partitioning have been reported in some fungi, the degree to which appears to be highly dependent on the species and experimental conditions. The closest reported examples are *Aspergillus niger* and *Aspergillus nidulans*, for which it has been reported that glucose uptake occurs more rapidly than fructose, whereas fructose-grown cultures demonstrate improved overall growth [46-49]. This evidence mirrors our findings for *F. venenatum*, in that fructose is primarily and efficiently utilised for tissue formation, while glucose is used primarily to support high respiration rates under the studied conditions. On the other hand, it has been found in *Penicillium janczewskii* that fructose supplementation resulted in fragile hyphae due to poor structural assimilation into cell walls, while the rate of fructose and glucose uptakes were similar [50], highlighting potential morphological considerations when comparing the two sugars. The underlying mechanistic explanations for such observations remain somewhat inconclusive in the literature for fungi and yeast.

Differences in glucose and fructose utilisation have been linked to alternative points of entry into central carbon metabolism, with fructose entering via direct phosphorylation rather than via glucose-6-phosphate isomerisation, although the causal mechanism as to why this would lead to higher biomass yield is unknown [46]. While uncommon, some strains contain fructose-specific transport systems, such as the microbial protein strain *Neurospora crassa* [51], which may result in differential carbon uptake and allocation between glucose and fructose. It is also possible that environmental factors played a role in this observed discrepancy. For example, it has been observed for *B. cinerea* that fructose uptake is highly pH dependent, decreasing dramatically above pH=6, suggestive of a unique proton-symport mechanism specific to uptake of fructose [52]. Overall, it is unclear from our findings the underlying causal mechanism to explain the differential utilisation of glucose and fructose by *F. venenatum*. Furthermore, while fructose supports high apparent biomass yield under the studied condition, its slow uptake rate and low specific growth rate may reduce conversion and mycoprotein production efficiencies at scale. The presence of a potential idiophase discussed previously, and reported negative effects on hyphal development in similar fungi warrant further investigations into morphogenesis, genetic expression, molecular carbon tracking and media optimisation to elucidate the differences between these sugars at the metabolic and system level. This may help to further understand their relative impacts on biomass composition and cellular health, and guide strategies to improve growth and utilisation rates for fructose-supported mycoprotein production.

## 4.4 Galactose and lactose: kinetic, transport and morphological limitations

For lactose and galactose cases, the slow uptake, adaptation and residence times indicated limiting factors unrelated to total carbon availability. Galactose demonstrated the longest modelled lag phase, while still following two sequential sigmoidal regimes. While lactose did not conform to the Gompertz framework, it exhibited slow linear growth, with sustained, monotonic utilisation of lactose occurring only after 96 h. In addition, although not formally presented, microscopic observation showed that the galactose culture still contained an observable population of conidia by the end of the cultivation, while the lactose culture was almost entirely constituted by conidia, with few, poorly developed hyphal fragments. This contrasted with all other conditions, where a thick mycelial mat was formed at the air-liquid interface, demonstrating a high degree of conidial germination and hyphal development. These morphological observations, the rate and onset of sustained substrate uptake, and the difference in growth dynamics between galactose and lactose serve to provide potential mechanistic insights into the response of *F. venenatum* to these related sugars.

One such explanation is that germination of *F. venenatum* conidia in lactose and galactose environments is poorly supported. There is reported evidence that for *Aspergillus niger* - another fungal strain important in bioproduction - that limited growth on galactose is attributable to a lack of activation of the key transport systems and germination signals [53-55]. On the other hand, it was found that actively growing mycelia demonstrated elevated galactokinase activities and expression of genes associated with the Leloir pathway (the metabolic sequence through which galactose enters central metabolism in fungi) [56-58]. This may explain the protracted lag duration observed for growth on galactose, in which conidia are slow to germinate, with the sustained growth periods associated with improved uptake of galactose by the developing hyphal networks.

On the other hand, as lactose was not observed to support sustained, exponential hyphal development during the experimental timeframe, further explanations are required to explain its unique phenotype. Firstly, in contrast to the other disaccharide studied (sucrose), lactose utilisation did not result in an accumulation of its monomers (glucose and galactose) in the extracellular medium. This suggests that lactose utilisation by *F. venenatum* relies on uptake of the intact disaccharide and subsequent intracellular hydrolysis. Genomic and physiological evidence of this mechanism has been found for several related fungal strains, such as *Fusarium graminaerum, Aspergillus nidulans, and Neurospora crassa*. These strains rely on lactose permease-mediated active transport of lactose into the cell, wherein it is hydrolysed by β-galactosidase enzymes into glucose and galactose [59]. Therefore, it is possible in the case of lactose, that a two-tier transport-kinetic limitation was imposed on its utilisation, with slow induction and/or capacity of lactose permeases at the transport level, and slow hydrolysis of lactose due to low activity of intracellular β-galactosidase at the metabolic level. When combined with, these limitations could severely inhibit sustained utilisation of lactose and lead to stunted hyphal outgrowth and accumulation of conidia. As the vegetative development filamentous morphology is critical to ensure high mycoprotein production rates and quality for food purposes, our findings under the current conditions appear to suggest that galactose and lactose-containing side streams (e.g. whey) are undesirable candidate feedstocks for scaled-up production.

To investigate the mechanistic galactose/lactose utilisation by *F. venenatum* further, future work could include measuring intracellular β-galactosidase activity, quantifying transporter gene expression, real-time microscopic tracking of morphological states, and supplementing cultures with low levels of glucose or galactose to assess induction effects on lactose catabolism.

## 4.5 Metabolic inheritance and non-additivity in diauxic systems

The results of the mixed-substrate experiments demonstrated that growth of *F. venenatum* on secondary carbon sources is strongly conditioned by the metabolic state established by the preceding utilisation of glucose. In general, glucose-xylose and glucose-galactose mixtures showed rapid glucose-supported growth followed by a slow diauxic transition to a stunted second growth phase, despite immediate uptake of the second sugar following glucose depletion. On its face, this seems to pose negative implications for the utilisation of mixed carbon substrates derived from waste streams containing glucose in mixture with secondary sugars. However, the results of these experiments should be interpreted in the context of potential uncontrolled environmental limitations arising from the first growth period in which glucose was utilised. This could explain the apparent non-additivity between xylose and galactose-only supported growth and their utilisation in mixture with glucose, as they were utilised when *F. venenatum* was at likely an entirely different metabolic and morphological state at the point of the diauxic transition. In this context, the behaviour of mixed-substrate systems cannot be interpreted as a simple superposition of single-substrate kinetics but instead reflects the legacy effects of the metabolic programme established during the initial glucose-supported growth phase.

This observation has also been found in yeast, where transition from fermentative growth on glucose, to respiratory reuptake of byproducts (including ethanol) can significantly alter the metabolic state through structural and functional mitochondrial shifts[60, 61]. Indeed, the diauxic shift lead to upregulation of mitochondrial and energy-generating pathways, in addition to genes encoding stress response products. Critically, downregulation of biosynthetic processes also occurred, leading to a severely reduced growth rate [61]. This reflects a shift from a high growth rate regime to a survival/slow growth regime in which carbon uptake is increasingly decoupled from biosynthetic growth and instead prioritised towards maintenance, redox balancing, and stress adaptation. This may be the case in our studied mixed substrate systems, where the metabolic state imposed by the uptake of ethanol following glucose depletion resulted in purely respiratory co-utilisation of the secondary sugar, leading to a slow and second growth phase, where the second sugar is utilised to lower biosynthesis efficiencies when compared with their single substrate counterparts. Beyond respiratory reprogramming, additional environmental feedback arising from the first growth phase may further reinforce this low-efficiency regime. It has been found in *S. fibuligera* that decreased pH from the production of acidic fermentation byproducts results in drastically reduced substrate utilisation efficiencies due to similar metabolic changes [62]. It is therefore possible that the sustained presence of exported organic acids (in particular pyruvate) during the second growth phase further exacerbated the low utilisation efficiencies of the secondary sugars.

On the other hand, despite xylose and galactose being poorly utilised for biosynthetic growth, their uptake commenced almost immediately following glucose depletion. This indicates that relief of glucose-mediated catabolite repression in *F. venenatum* occurs on a short timescale, enabling rapid uptake of secondary carbon sources. Importantly, this rapid onset appears to be decoupled from efficient biomass formation, suggesting that regulatory control of substrate utilisation and the physiological capacity for biosynthesis were governed by distinct constraints. Similar behaviour has been reported in ectomycorrhizal fungi, where glucose addition enhanced the uptake of secondary carbon sources without necessarily supporting proportional increase in yield [63].

Overall, the findings of the mixed-substrate studies strongly indicate that utilisation efficiencies in diauxic systems are a function of both substrates contained. Therefore, the kinetic screening results of alternative substrates *in silo* should not necessarily be taken as representative of their kinetics in mixed systems representative of global side streams. This poses a potential design challenge in the utilisation of these streams, in that the state induced by overflow on glucose should be minimised at scale to improve yields associated with secondary carbon sources. Therefore, mixtures should be further investigated in scaled-up screening experiments, wherein strategies could be employed to reduce the extent of overflow metabolism and the deleterious effects of its byproducts to optimise biomass yields. For example, the extent of fermentative metabolism from glucose may be reduced by improved aeration, varying the concentration of glucose, and adapting feed strategies, while pH control may mitigate the negative effects of increasing acidity on the bio-formation efficiencies associated with the secondary substrate [64].

## 4.6 Expired functional drink as a candidate waste feedstock and citric acid effects

The expired functional drink experiment represents an escalation from defined synthetic systems to a compositionally complex, real-world feedstocks. The growth trajectory observed for the expired soft drink deviated substantially from that of the synthetic control. Rather than the more gradual and balanced progression observed in the synthetic condition, the expired drink supported the highest estimated maximum specific growth rate in this study, the shortest times to key growth milestones ($T_{10}$, $T_{50}$ and $T_{90}$) and the highest final biomass titre. Furthermore, the extent of overflow was substantially lower in the case of the expired soft drink, with higher rates of reoxidation following depletion of sugars. The windowed apparent biomass yield from sugars was also substantially higher than that of the synthetic control, and the extent of co-utilisation (between glucose and fructose) was also improved compared to the synthetic control and all other dual sugar systems examined, which tended to exhibit strict sequential utilisation behaviour. On the basis of these metrics alone, functional drink waste would appear to be an exceptionally promising feedstock for mycoprotein production when compared with synthetic media systems.

The distinct kinetic behaviours exhibited by both the real and control systems motivated experimental investigation into the effect of citric acid addition to media containing the same substrates studied in Sections 3.1 and 3.2, the results of which are presented fully in Supplementary Information 6 and 7 for single and dual substrates respectively. In brief, the results demonstrated a general trend of reduced growth rates and higher biomass yields across substrates in comparison to the unmodified media conditions, in addition to sustained and improved late-stage growth in systems exhibiting sequential uptake of multiple substrates.

In particular, the prolonged early-stage growth observed across the single and dual substrate conditions, reflected by consistently increased $T_{10}$ and $T_{50}$, suggests that citric acid may impose an initial growth constraint prior to onset of rapid and expansive growth. A plausible contributing factor could be chelation of trace elements essential for biosynthesis by high concentrations of citric acid. Citric acid is already included at low levels in the original media formulation to maintain solubility and stability of iron species $Fe^{2+}$ and $Fe^{3}$. However, supplementation at 7g·L$^{-1}$ may periodically reduce the bioavailability of divalent cations required for germination and early hyphal growth, a known inhibitory effect of organic acids on fungi such as *Candida albicans* [65, 66]. Importantly, these effects appear transient, as cultures subsequently recover and exhibit enhanced late-stage growth and high overall biomass yield, indicating that early inhibition does not constrain overall biosynthetic capacity, an observation characteristic of trace-element limited cultures [67].

Early physiological studies of filamentous fungi reported that citric acid could be tolerated at high concentrations and influence growth behaviour without acting as a primary or immediately assimilated carbon source, with utilisation often occurring only after other carbon sources were exhausted [68]. Indeed, the ability to successfully propagate in organic acid environments is critical to many citrus crop-degrading fungal strains such as *Fusarium graminaerum* [69]. More specific phenotypic precedents align well with our findings. For example, the observation that citric acid delays the time to maximum biomass while increasing the final biomass titre is consistent with reports for *Aspergillus glaucus*, where addition of 1 to 4 g·L$^{-1}$ citric acid to glucose cultures resulted in up to 15.7% higher biomass despite longer times to maximum CDW compared with a glucose-only control [70]. In the same study, citric acid also enhanced glucose uptake by more than 10%, indicating improved metabolic efficiency rather than simply inhibiting early growth.

Evidence for a regulatory metabolic influence of citric acid has been demonstrated in several fungal species. For instance, strong support for a potential regulatory role of citric acid is provided by recent transcriptomic and proteomic analysis of *F. graminearum*, which demonstrated that citric acid supplementation promoted fungal cellular growth and differentially repressed secondary metabolite production, particularly at higher supplied concentrations [71]. Notably, citric acid induced upregulation of carbon metabolism enzymes including phosphoenolpyruvate carboxykinase (PEPCK), a key enzyme linking the tricarboxylic acid (TCA) cycle with biosynthetic and gluconeogenic pathways, even when sugars remained the dominant carbon source [72]. Furthermore, upregulation of pathways directly associated with gluconeogenesis and biosynthesis precursor supply (e.g. glycerol kinase) was observed in the presence of citric acid. These findings suggest that citric acid can alter carbon metabolism independently of its direct assimilation, enhancing metabolic flexibility particularly in support of biosynthetic capacity. The regulatory influence of citric acid may therefore explain our observations of increased sugar uptake rates corresponding with higher sugar-biomass yield efficiencies and sustained late-stage growth in the presence of citric acid for *F. venenatum* here.

In *Aspergillus glaucus*, elevated citric acid concentrations were also associated with a shift in biosynthesis strategy during late-stage growth, with enhanced production of the polyketide aspergiolide A as residual glucose concentration declined [70]. While secondary metabolite formation was not assessed for *F. venenatum* here, these findings support a broader interpretation that citric acid can alter intracellular carbon allocation processes to enable alternative biosynthetic outcomes at different stages of growth.

Supplementation of organic acids have also been shown to improve metabolic efficiencies in a variety of microbial systems by increasing TCA-linked redox capacity. In denitrifying cultures, citrate was found to promote the activity of isocitrate dehydrogenase, improving oxidative phosphorylation efficiency through accelerated NAD+ generation [73]. Similarly, studies of acetic acid utilisation in *Fusarium oxysporum* demonstrate that organic acids can have a significant effect on central metabolic processes. In mixed glucose-acetate cultures, acetate was consumed only after glucose depletion, but its presence was associated with elevated TCA cycle intermediates, induction of isocitrate lyase and improved growth rates and biomass yields within a specific concentration range, but yielded biomass with substantially altered amino acid profiles. Notably, high acetate concentrations (~ $8 g \cdot L^{-1}$) also increased the lag phase duration, mirroring the delayed early growth observed for *F. venenatum* in this study with $7 g \cdot L^{-1}$ citric acid [74]. Therefore, it is postulated that citric acid supplementation may enhance anaplerosis (resupply of TCA intermediates to support sustained biosynthesis) and TCA-linked redox capacity in *F. venenatum*, enabling more efficient carbon conversion to biomass particularly during periods of high respiratory demand. This interpretation is consistent with the observed reduction in ethanol yields, indicating a reduced reliance on fermentative overflow products as redox sinks during growth. Furthermore, the complete absence of measured extracellular pyruvate in the citric acid amended systems may be a further indicator of relieved metabolic stress, as pyruvate export has been shown to be a key indicator of cellular stress control against accumulated reactive oxygen species (ROS) during respiration [75]. At the same time, increased substrate uptake rates coupled with improved biomass yield efficiencies suggest more effective coupling between carbon catabolism and biosynthesis. The ability of *F. venenatum* to sustain substantial growth on secondary sugars and oxidative substrates further suggests that central metabolism, particularly TCA-linked respiration, remains sufficiently robust to support uninterrupted biosynthesis during mid-to-late-stage growth under citric acid supplementation.

Organic acids including citric acid have also been shown to influence microbial growth efficiency not only through carbon metabolism modulation effects but also via acid-base homeostasis and metabolic stabilisation. Citric acid uptake in some fungi has been reported to occur via proton-symport proteins in the plasma membrane, with strong dependence on pH conditions where citric acid is present in its dissociated form (citrate) [76-79]. This may suggest that the citric acid uptake observed during late-stage growth have an alkalising effect on the growth medium. Furthermore, the presence of citrate has been associated in some species to stimulate growth and increase the specific sugar consumption rate, due to reduced energetic cost of pH regulation during periods of high respiratory activity and acid stress [79]. This could be an alternative or additional potential explanation for the observations of more efficient sugar utilisation in all conditions but should be further investigated experimentally through online pH measurements in future works.

While citric acid was determined to have significant singular effects on the kinetics of *F. venenatum* growth, the real expired functional drink sample also exhibited a strong temporal misalignment between biomass accumulation and measured substrate utilisation, including an apparent late-stage biomass yield that could not be reconciled with citric acid uptake alone. This indicates that citric acid, while influential, cannot fully account for the observed dynamics, and instead likely acts in concert with additional, unquantified components of the feedstock. Given that citric acid was the only non-sugar carbon source detected by the applied HPLC protocol, the strong divergence from the synthetic citric acid condition suggests that other constituents present in functional drink exert interacting, and potentially non-linear, effects on growth and carbon utilisation. This highlights potential limitations of the current analytical chemistry approach adopted in this work, which was adequate only for the quantification of major monosaccharides, sucrose and citric acid. As a result, additional carbon sources likely to be present in the functional drink, including minor organic acids such as malic, oxalic acid, free amino acids, proteins, and other soluble components, may have contributed to sustained growth but remained unaccounted for in the mass balance [7, 17, 80, 81]. Indeed, manufacturer information for the drink used indicates non-sugar carbohydrates and fibre can typically constitute ~ 4 $g \cdot L^{-1}$ and protein at ~2 $g \cdot L^{-1}$, which may be utilised as non-negligible sources of carbon and nitrogen for growth. Furthermore, the manufacturer states the presence of ascorbic acid (vitamin C) at ~ 0.3 g·$L^{-1}$, which has been shown in yeast and fungal systems to relieve oxidative and pH-induced cell stress factors [82-84]. Such minor compounds could plausibly support more rapid exponential growth both during and following periods of high respiratory activity by reducing the accumulation of intracellular reactive oxygen species (ROS), thereby mitigating their associated inhibitory metabolic effects. Although this mechanism was not directly assessed here, it provides an avenue for further investigation into the media matrix effects on *F. venenatum* growth in compositionally complex waste feedstocks. Future work incorporating total organic carbon (TOC) and targeted extracellular metabolite quantification through pairing of liquid chromatography with mass spectrometry (LC-MS) may help enable contributions of minor and major unresolved compounds to be fully resolved in non-synthetic, complex environments.

In addition, an interesting qualitative difference was observed in biomass appearance, with material recovered from the expired drink exhibiting a distinct pink-orange coloration relative to the white biomass obtained under synthetic conditions (Supplementary Information 8). This may reflect uptake of pigments such as carotenoids naturally present in functional drink, which have been shown to have a similar antioxidant effect on ROS as vitamin C [84]. Alternatively, this colour change may reflect a shift in biosynthetic or morphological state during late-stage growth through upregulated production of pigments as secondary metabolites, indicating a shift in biosynthetic strategy. The first plausible explanation of uptake from juice may further explain the increased growth rates and respiratory efficiency, while the latter may be partially explanatory of the yield discrepancy during the phase of citric acid utilisation in expired juice.

Without time-series biocompositional analysis, it was assumed throughout the entire growth period that CDW consisted solely of cellular biomass with previously determined elemental composition of industry-scale, glucose-supported mycoprotein product. However, it has been shown in many works that the media environment and growth-substrate identity can strongly influence the biosynthetic strategy of fungal cultures resulting in varying compositional outcomes. For instance, when supplemented with citric acid, *A. niger* cultures supplemented showed an increase in fatty acid content of biomass, while in *Fusarium culmorum* mycelial mats, citrate enhanced the nitrogen content of the biomass through increased uptake of ammonium nitrogen [85, 86]. Such effects have been associated with alternative biosynthetic routing through glyoxylate shunt metabolism of non-fermentable carbon sources (e.g. ethanol) facilitated by citric acid [87]. It is therefore essential in future work to quantify the temporal evolution of biomass composition, including protein and amino-acid content, lipid fractions, and secondary metabolites such as pigments alongside growth and substrate utilisation. Such compositional resolution is critical for assessing downstream processing requirements and product quality at scale (i.e. nutritional profiles, protein content, colour and texture), while secondary metabolites may include those which would jeopardise product safety (e.g. mycotoxins). These analyses may also help to resolve observed regions of unexplained decoupling between measured carbon uptake and assumed cellular biomass accumulation, which would allow for accurate determination of the relative carbon conversion to biomass solely constituting mycoprotein.

Furthermore, from a process perspective, the strong catabolite repression of citric acid observed raises additional questions regarding how such dynamics would translate to continuous operation at scale. Maintaining sufficient co-consumption of substrates while avoiding accumulation of strongly repressed organic acids (which may become increasingly inhibitory at higher concentrations) may impose trade-offs between feedstock conversion efficiency and volumetric productivity. Addressing these questions will require controlled continuous bioreactor experiments coupled with compositional analysis of biomass to assess both performance and suitability of expired functional drink and other waste-derived feedstocks for food-grade applications.

Overall, from an application perspective, this feedstock exhibited exceptional performance, with growth rates and biomass titres exceeding all synthetic systems examined in this study, highlighting the potential of *F. venenatum* to effectively utilise complex carbon resources. From a methodological perspective, this study represents the first application of high-resolution, time-series analysis combined with phenomenological growth modelling to *F. venenatum* cultivated on a real waste feedstock. Importantly, this framework enables deviations between biomass formation and measured substrate utilisation to be identified and localised in time to provide a structured basis for targeted follow-up experiments where synthetic media systems alone are insufficient to explain observed behaviour.

# 5 Conclusion

This work extended and refined the high-throughput batch microlitre experimental workflow developed previously to systematically evaluate growth, substrate utilisation and extracellular byproduct dynamics by *F. venenatum* across a broad range of synthetic sugars and their mixtures, representative of those present in low-cost and abundant global side streams in addition to a real feedstock in expired functional drink.

Across single substrates, several interesting trends emerged. It was found from unified Gompertz analysis that the growth phenotype was highly substrate-specific, and that a single sigmoidal regime was not sufficient to describe all conditions. These models, while relying solely on the measurement of biomass and limited in their mechanistic insights alone, provided a framework for co-analysis with substrate and byproduct trends determined through analytical chemistry methods. Indeed, it was found broadly that the Gompertz predicted growth dynamics and metrics aligned well with the data-derived substrate and byproduct event timings, rates and yields, supporting biological interpretability and the robustness of this modelling approach.

It was found that substrate identity has a significant impact on the resulting phenotypes, with consistently observed trade-offs between critical metrics determinative of performance in bioprocesses. Rapidly consumed sugars such as glucose and sucrose supported high growth rates at the expense of biomass yield, while sugars consumed at a slower rate (xylose, fructose, and galactose, lactose) favoured reduced growth rates. However, in the case of the latter group, the impact on biomass yield was more complex and varied, indicative of other substrate-specific effects related to morphogenesis and transport/pathway induction effects.

Broadening the analytical scope of the methodology to include byproduct dynamics proved highly valuable in providing plausible interpretations for the observed trends. Apparent growth inefficiencies could be explainable by the substantial diversion of carbon to fermentative products, associated with overflow metabolism due to high carbon flux, and/or fermentative growth resulting from a transition from to aerobic to anaerobic conditions as the mycelial cultures develop and respiratory oxygen demand increases.

The mixed sugar systems explored in this work are of critical relevance to real waste streams containing a diversity of fermentable substrates. Results showed that single substrate conditions for xylose, galactose, and fructose were not predictive of their utilisation efficiencies in mixture with glucose. While uptake of secondary sugars commenced quickly following glucose depletion, they were utilised to lower biomass growth efficiencies alongside reoxidation of fermentative byproducts. Collectively, these findings indicate that secondary carbon utilisation in diauxic growth of *F. venenatum* is strongly conditioned by the metabolic and environmental state established during the initial growth period. This has important implications for the design of scaled-up mycoprotein production from compositionally complex feedstocks, for example, through the use of tailored feeding strategies and reactor configurations to minimise byproduct accumulation and maximise biomass yields.

Cultivation of *F. venenatum* on expired functional drink resulted in the fastest growth, highest biomass titre, and most efficient apparent carbon conversion observed across all conditions studied in this thesis. Relative to a compositionally matched synthetic control, the real feedstock supported substantially higher sugar uptake rates, reduced fermentative overflow, enhanced byproduct reassimilation, and improved biosynthesis from substrate utilisation and biomass accumulation during early and mid-growth phases. These results confirm that beverage-industry waste streams can represent highly effective feedstocks for microbial protein production and that performance inferred solely from sugar composition may substantially underestimate true growth potentials in real mixtures.

The findings of this work demonstrate the significant value of the novel high throughput screening framework as a tool to guide future experimental designs and methodological enhancements, such as enabling targeted investigations of substrate-specific and system constraints. These may include spectrophotometric quantification of dissolved oxygen and pH profiles to accurately characterise the environmental state of the systems, with the potential for real-time feedback automated control through addition of a pH neutraliser to plate-wells and orbital shaking regimes to improve aeration. Furthermore, biomass measurements beyond $OD_{600}$ and CDW could reveal effects on growth development of *F. venenatum* in response to different substrates and environmental shifts over time. These could include morphological development analyses through integration of real-time microscopic imaging, and time-series profiling of compositional changes in biomass, for example using semi-quantitative, high throughput, label-free spectroscopic approaches (e.g. ATR-FTIR).

In conclusion, this work showed that the growth performance of *F. venenatum* varied significantly across industrially relevant sugars and real feedstock samples. Critically, byproduct dynamics served to elucidate differences in growth dynamics and substrate utilisation between substrate types which may not have been identifiable without their measurement. By integrating high throughput batch microplate screening with comparative, data-driven analyses of growth, substrate utilisation and byproduct dynamics, this work provides a strong foundation for evaluating the suitability of diverse carbon substrates and their mixtures for mycoprotein production and establishes clear guidelines for targeted experimental analyses and methodological enhancements. Overall, the insights obtained from this study move beyond substrate screening towards a framework for informed decision-making in the bioprocess development and design pipeline at an early stage, for accelerated scale-up of mycoprotein production from complex waste carbon resources.

## Author contributions

M.B. designed the study, performed research and data analyses. M.B., N.S., Y.H., T.V. performed experimental work. Y.H. contributed to experimental protocol development. M.B. and M.G. drafted the manuscript sections. All authors contributed to the final paper revision and approved the final manuscript.

## Conflicts of interest

There are no conflicts to declare.

## Acknowledgements

M.B. and M.G. would like to acknowledge the funding support provided by EPSRC DTP Industrial CASE Studentship (project reference 2609294) and Marlow Foods.